\documentclass[envcountsect,runningheads]{llncs}

\usepackage[T1]{fontenc}%
\usepackage{graphicx}
\usepackage{amsmath}
\usepackage{amssymb}
\usepackage{amsfonts}
\usepackage{ascmac}
\usepackage{booktabs}
\usepackage{multirow}
\usepackage{xspace}
\usepackage{color}
\usepackage{colortbl}
\usepackage{wrapfig}
\usepackage{cite}
\usepackage{stmaryrd}
\usepackage{bbding}
\usepackage{comment}
\usepackage[long]{optidef}
\usepackage[]{todonotes}
\usepackage{enumitem}
\usepackage[skip=0pt]{subcaption}

\usepackage{lineno} 

\usepackage{pifont}

\newcommand{\AP}{\ensuremath{\mathrm{AP}}}

\DeclareMathOperator{\X}{\mathbf X}
\DeclareMathOperator{\XX}{\mathbb X}
\DeclareMathOperator{\F}{\mathbf F}
\DeclareMathOperator{\FF}{\mathbb F}
\DeclareMathOperator{\G}{\mathbf G}
\DeclareMathOperator{\GG}{\mathbb G}
\newcommand{\U}{\ensuremath\mathbin\mathbf{U}}
\newcommand{\R}{\ensuremath\mathbin\mathbf{R}}
\DeclareMathOperator{\Y}{\mathbf Y}
\newcommand{\Oop}{\ensuremath\operatorname{\mathbf{O}}}
\newcommand{\Hop}{\ensuremath\operatorname{\mathbf{H}}}
\newcommand{\Sop}{\ensuremath\mathbin\mathbf{S}}
\newcommand{\T}{\ensuremath\mathbin\mathbf{T}}
\newcommand{\req}{\ensuremath\mathit{req}}
\newcommand{\resp}{\ensuremath\mathit{resp}}
\newcommand{\call}{\ensuremath\mathit{call}}
\newcommand{\internal}{\ensuremath\mathit{int}}
\newcommand{\ret}{\ensuremath\mathit{ret}}
\newcommand{\RR}[1]{\ensuremath\mathsf{RR}{#1}}
\newcommand{\baseRR}{\ensuremath\mathsf{baseRR}}

\renewcommand{\succ}{\ensuremath\operatorname{\mathrm{succ}}}

\newcommand{\myparagraph}[1]{\noindent\textbf{#1}\quad}

\newcommand{\place}{\underline{\phantom{n}}\,}
\newcommand{\LRRthree}{L_{\mathrm{RR3}}}
\newcommand{\LRRfour}{L_{\mathrm{RR4}}}

\usepackage[amsmath]{ntheorem}
\usepackage{cleveref}

\theoremnumbering{arabic}
\theoremseparator{.}
\theorembodyfont{\itshape}
\newtheorem{mytheorem}{Theorem}[section]

\newtheorem{myproposition}[mytheorem]{Proposition}

\theorembodyfont{\rmfamily}

\newtheorem{myremark}[mytheorem]{Remark}

\newtheorem{mydefinition}[mytheorem]{Definition}
\newtheorem{myassumption}[mytheorem]{Assumption}

\newtheorem*{myproof}{Proof}

\creflabelformat{enumi}{#2(#1)#3}
\crefname{mytheorem}{Thm.}{Thms}
\crefname{mydefinition}{Def.}{Defs}
\crefname{myproposition}{Prop.}{Props}
\crefname{myremark}{Rem.}{Remarks}
\crefname{mylemma}{Lem.}{Lemmas}
\crefname{myproof}{Proof.}{Proofs}
\crefname{myproblem}{Prob.}{Proobs}
\crefname{myassumption}{Assum.}{Assumptions}
\crefname{myexample}{Ex.}{Exs}
\crefname{appendix}{Appendix}{Appendixes}
\crefname{algorithm}{Alg.}{Algs}
\crefformat{section}{{\S}#2#1#3}
\crefname{figure}{Fig.}{Figs}
\Crefname{equation}{}{}
\crefname{table}{Table}{Tables}

\usepackage{tikz}
\usepackage{tikz-cd} 
\newcommand*\circled[1]{\tikz[baseline=(char.base)]{
            \node[shape=circle,draw,inner sep=3pt] (char) {#1};}}
\newcommand*\doublecircled[1]{\tikz[baseline=(char.base)]{
			\node[shape=circle,draw,inner sep=2pt] (char) {#1};\node[shape=circle,draw,inner sep=3pt] (char) {#1};}}

\specialcomment{auxproof}
{\mbox{}\newline\textbf{BEGIN\@: AUX-PROOF}\dotfill\newline}
{\mbox{}\newline\textbf{END\@: AUX-PROOF}\dotfill\newline}
\excludecomment{auxproof}
\ifdefined\VersionLong
    \includecomment{LongVersion}
    \excludecomment{ShortVersion}
\else
    \excludecomment{LongVersion}
    \includecomment{ShortVersion}
\fi

\ifdefined\VersionFinal%
  \newcommand{\FinalVersion}[1]{#1}
  \newcommand{\ArxivVersion}[1]{}
  \newcommand{\FinalOrArxivVersion}[2]{#1}
\else
  \newcommand{\FinalVersion}[1]{}
  \newcommand{\ArxivVersion}[1]{#1}
  \newcommand{\FinalOrArxivVersion}[2]{#2}
\fi
\usepackage{comment}

\ifdefined\VersionFinal%
  \excludecomment{ArxivBlock}
\else
  \includecomment{ArxivBlock}
\fi

\begin{document}
\title{A Variety of Request-Response Specifications\thanks{The authors are partially supported by  the ASPIRE grant (No.\ JPMJAP2301) \& CREST ZT-IoT Project (No.\ JPMJCR21M3), JST.}}
\titlerunning{A Variety of Request-Response Specifications}
\author{
Daichi Aiba\inst{1}
\and
Masaki Waga\inst{2,1}
\and
Hiroya Fujinami\inst{1,5}
\and
Koko Muroya\inst{3}
\and
Shutaro Ouchi\inst{4}
\and
Naoki Ueda\inst{4}
\and
Yosuke Yokoyama\inst{4}
\and 
Yuta Wada\inst{4}
\and
Ichiro Hasuo\inst{1,5,6}
}
\authorrunning{
D.\ Aiba  et al.}
\institute{
National Institute of Informatics, Tokyo, Japan \\
\email{\{daichi-aiba,makenowjust,hasuo\}@nii.ac.jp}
\and
Kyoto University, Kyoto, Japan\\
\email{mwaga@fos.kuis.kyoto-u.ac.jp}
\and
Ochanomizu University, Tokyo, Japan\\
\email{kmuroya@is.ocha.ac.jp}
\and
Mitsubishi Electric Corporation, Kamakura, Japan\\
\email{\{Ouchi.Shutaro@ab,Ueda.Naoki@cw,Yokoyama.Yosuke@ea,Wada.Yuta@ab\}.}\\
\email{mitsubishielectric.co.jp}
\and
SOKENDAI, Tokyo, Japan
\and
Imiron Co., Ltd., Tokyo, Japan
}
\maketitle
%
%
\vspace{.1em}
\noindent
\begin{minipage}{\textwidth}
\begin{abstract}
 We find, motivated by real-world applications, that the well-known \emph{request-response specification} comes with multiple variations, and that these variations should be distinguished. As the first main contribution, we introduce a classification of those variations into six types, and present it as a decision tree, where a user is led to the type that is suited for their application by answering a couple of questions. Our second main contribution is the formalization of those six types in various formalisms such as grammars, temporal logics,  and automata; here, two types out of the six are non-regular specifications and their formalization requires extended formalisms.  We also present an overview of  tools for monitoring these specifications to cater for practitioners' needs.  Through these contributions, we address an  issue in formal specification that is practically important but has been scientifically somewhat overlooked.
\end{abstract}
\end{minipage}

\section{Introduction}\label{sec:intro}
\emph{Formal specification} refers to the process of pinning down requirements for a target system in a rigorous, mathematical and formal language. It is a prerequisite of \emph{formal verification}, where one aims to prove $\mathcal{M}\models \varphi$ ($\mathcal{M}$ \emph{satisfies} $\varphi$), given a formal model $\mathcal{M}$ of a system and a formal specification $\varphi$. 

However, with the recent rise of diverse computing paradigms that make white-box models $\mathcal{M}$ expensive or impossible to get, the importance of formal specification itself is increasingly recognized. 
Examples of such paradigms include cyber-physical systems (whose physical components are hard to model) and machine learning systems (where neural networks are large numerical objects and hard to model logically). For those systems, formal verification is often too expensive, and one would turn to \emph{modelless}, \emph{spec-only} techniques such as monitoring and property-based testing. Although these techniques do not give the guarantee level of formal verification, they can automate various quality assurance tasks on a solid basis of mathematical representations of requirements, and thus attract attention both from academia and industry. See e.g.,~\cite{SanchezSABBCFFK19,BartocciDDFMNS18,MenghiABEFFGK0P23}.

One class of formal specifications that appear often in many application domains is \emph{request-response specifications} (\emph{req-resp specs}). They require that, whenever a request $\req$ is issued, it has to be responded by $\resp$. Examples include error/anomaly handling, server-client interaction, commercial transactions, communication protocols, etc.\ (some concrete examples are in \cref{subsec:examples}). Req-resp specs are often expressed by LTL formulas of the form $\G(\req\to\F\resp)$---at any moment ($\G$), if $\req$ occurs, then eventually ($\F$), $\resp$ should occur.

 This work is motivated by the observation that the simple spec $\G(\req\to\F\resp)$  \emph{may not adequately express the designer's intention}. Consider, for example, a word $\req\,\req\,\resp$ of $\req$'s and $\resp$'s. It does satisfy the formula $\G(\req\to\F\resp)$, but in case we count the number of pending $\req$'s (e.g., when $\req$ stands for a coin put in a vending machine), one $\resp$ is not enough to resolve two $\req$'s. Therefore we have to consider multiple types of req-resp specs, and moreover, to clarify the criteria for choosing a type that is suited for each specific application.

In this paper, we make the following contributions, 
through which we address a problem that has often discouraged industrial adoption of formal technologies.

\vspace{.2em}
\noindent
\begin{minipage}{\textwidth}
 \begin{itemize}
 \item We present a classification of req-resp specs into six types (RR1--6, \cref{sec:clTree}). The classification is organized in a decision tree (\cref{fig:clTree}), where a user is led to a suitable type by answering a couple of questions (criteria C1--C3).
 \item We formalize the six types in three formalisms, namely grammars, temporal logic formulas, and automata (\cref{sec:formalization}). Two types are  non-regular (RR3,4); for them, we show which extensions of the common formalisms can be used. There are some non-trivial observations here, too, such as \cref{thm:suffixcount}.
 \item We evaluate our classification  (\cref{sec:assessments}), assessing if its level of distinction is appropriate and if its decision tree presentation is sensible. The examples we present for the six types in \cref{subsec:examples} help our case, too.
 \item We present an overview of monitoring tools for req-resp specs. This will cater for practitioners' needs of monitoring data at their hand.
 \end{itemize}
\end{minipage}




\section{Preliminaries}\label{sec:prelim}

% ...


We introduce some formalisms for temporal specifications.
Many of them are  standard;
for them, our purpose here is to fix notations. 
Others are less known;
for them, we provide a brief introduction.

\myparagraph{Notations} 
Fix a finite alphabet $\Sigma$. 
For a finite word $w= w_0w_1\dots w_{n-1}\in\Sigma^*$, the $i$-th element is denoted by $w_i\in \Sigma$. 
The suffix $w_{i}w_{i+1}\dots w_{n-1}$ of $w$ is $w_{i\le}$; the prefix $w_{0}w_{1}\dots w_{i}$ is  $w_{\leq i}$.  $\#a$ is the number of occurrences of $a\in\Sigma$ in a given word. The empty word is $\varepsilon$. $\AP$ is a set of \emph{atomic propositions}. 
A partial function is denoted by $f\colon X {\,\rightharpoonup\,} Y$ with $\rightharpoonup$, where $f(x)=\bot$ denotes $f(x)$ is undefined.

\subsection{Regular Expressions and Context-Free Grammars}\label{subsec:regexpCfg}

\emph{Regular expressions} over an alphabet $\Sigma$, as usual, are defined by the following abstract grammar:
\begin{math}
 r \; ::= \; 0 \mid \sigma \mid 1 \mid (r|r) \mid r\cdot r \mid r^{\ast} 
\end{math}.
Here $0$ is for the empty language $\emptyset$,
 $\sigma$ denotes the singleton language $\{\sigma\}$ (with $\sigma\in\Sigma$), $1$ denotes  $\{\varepsilon\}$, $r|r$ denotes alternation, $r\cdot r$ denotes concatenation, and $r^{\ast}$ denotes repetition. We often drop $\cdot$ in $r\cdot r$, and $r^{+}$ is short for $r\cdot r^{\ast}$.

A \emph{context-free grammar} is a quadruple $(N,\Sigma,P,S)$, where $N$ is a finite set of nonterminals, $\Sigma$ is a finite set of terminals, $P$ is a set of production rules, and $S\in N$ is a start variable. A production rule is of the form $\alpha\to\beta$, where $\alpha\in N$ is a nonterminal and $\beta\in(N|\Sigma)^*$ is a word over nonterminals and terminals.

\subsection{Linear Temporal Logic (LTL)}\label{subsec:LTL}
The (usual, future-time) \emph{LTL} has propositional formulas extended with temporal operators  $\X$ (next), $\F$ (future), $\G$ (globally), $\U$ (until), and $\R$ (release). We use LTL extended with \emph{past-time} temporal operators, namely  $\Y$ (yesterday), $\Oop$ (once), $\Hop$ (historically), $\Sop$ (since), and $\T$ (trigger). Past-time operators have been found useful for expressing some specs---also in the current paper, 
see \cref{subsec:LTLformula}. 
 

\vspace{.2em}
\noindent
\begin{minipage}{\textwidth}
\begin{mydefinition}[syntax of LTL]
	We define LTL formulas as follows: \\
\begin{math}
 \varphi\;::=\;\bot\mid p\mid\lnot\varphi\mid\varphi\land\varphi\mid\X\varphi\mid\Y\varphi\mid\varphi\U\varphi\mid\varphi\Sop\varphi
\end{math} (here $p\in \mathrm{AP}$). 
	In addition, we use some abbreviations: 
	$\top=\lnot\bot$, $\varphi\lor\psi=\lnot(\lnot \varphi\land\lnot\psi)$, $\varphi\R\psi=\lnot(\lnot\varphi\U\lnot\psi)$, $\varphi\T\psi=\lnot(\lnot\varphi\Sop\lnot\psi)$, $\F\varphi=\top\U\varphi$, $\G\varphi=\bot\R\varphi$, $\Oop\varphi=\top\Sop\varphi$, $\Hop\varphi=\bot\T\varphi$.
\end{mydefinition}
\end{minipage}

\vspace{.3em}
\noindent
The semantics is defined as usual over finite words. See 
\FinalOrArxivVersion{\cite[Appendix~A.1]{Aiba_ICTAC2025_arxiv_extended_version}}{\cref{appendix:LTLsem}}.

\subsection{CaRet}\label{subsec:caret}
As announced in \cref{sec:intro}, two types of req-resp specs (namely RR3,4) are not regular, and thus  are not expressible in LTL.  We use an extension of LTL, namely the \emph{temporal logic of calls and returns (CaRet)}~\cite{AlurEM04}.
Here we introduce CaRet as originally presented in~\cite{AlurEM04}; later in \cref{subsec:LTLformula} we introduce and use its slight variation.

In CaRet, they consider two kinds of correspondences between \emph{calls} and \emph{returns}: the \emph{matching-return} correspondence $R$, and \emph{innermost-call} $Q$. Intuitively, $R$ maps $i$ to \emph{the first unmatched return} after $i$, and $Q$ maps $i$ to \emph{the most recent pending call} before $i$.
 Using them,   three  successor functions are  introduced.

\begin{mydefinition}\label{def:matching-ret-innermost-call}
Let $\widehat\Sigma=\Sigma\times\{\call,\internal,\ret\}$. For $\alpha\in\Sigma$, we call the symbols of the form $(\alpha,\call),(\alpha,\internal),(\alpha,\ret)\in\widehat\Sigma$, \emph{calls}, \emph{internals}, and \emph{returns}, respectively. We thus have three disjoint finite alphabets: $\Sigma_c$ is a finite set of calls, $\Sigma_{\internal}$ is a finite set of internals, and $\Sigma_r$ is a finite set of returns, with $\widehat\Sigma=\Sigma_c\cup\Sigma_{\internal}\cup\Sigma_r$.

	Let $\sigma$ be a finite word on $\widehat\Sigma$.
 The \emph{matching-return} correspondence $R_\sigma:\mathbb{N}\rightharpoonup\mathbb{N}$ on $\sigma$ is a partial function 
defined as follows.
(Case 1) If there is $j$ such that 1) $j>i$, 2) $\sigma_{j}$ is a return, and 3) the numbers of calls and returns from $i+1$ to $j-1$ are equal, then $R_\sigma(i)$ is the smallest such $j$.  (Case 2) Otherwise, $R_\sigma(i)=\bot$.

		 The \emph{innermost-call}  $Q_\sigma:\mathbb{N}\rightharpoonup\mathbb{N}$ on $\sigma$ is a partial function 
defined as follows. (Case 1)  If there is $j$ such that 1) $j<i$, 2) $\sigma_{j}$ is a call, and 3)  $R_\sigma(j)> i$ or $R_\sigma(j)=\bot$, then $Q_\sigma(i)$ is the greatest such $j$. (Case 2) Otherwise, $Q_\sigma(i)=\bot$.

If $\sigma_{i}$ is a call, 
 $\sigma_{j}$ is a return, and $R_\sigma(i)=j$, then we say that $\sigma_j$ is \emph{the matching return} for $\sigma_i$. 
Similarly, if $\sigma_{i}$ is a return, 
 $\sigma_{j}$ is a call, and $Q_\sigma(i)=j$, then we say that $\sigma_j$ is \emph{the innermost call} for $\sigma_i$.
For a quick intuition, see \cref{fig:matching_resp_RR3_and_RR4}.
\end{mydefinition}

\noindent
\begin{minipage}{\textwidth}
\begin{mydefinition}
Let $\sigma$ be a finite word on $\widehat\Sigma$. Three \emph{successor functions}
\linebreak
\begin{math}
 \succ^g, \succ^a, \succ^- \colon \mathbb{N}\rightharpoonup\mathbb{N}
\end{math} 
are defined as follows.


The \emph{global-successor} function $\succ_\sigma^g$ is defined by $\succ_\sigma^g(i) = i + 1$. 


The \emph{abstract-successor} function $\succ_\sigma^a$ is defined as follows:
$\succ_\sigma^a(i)=R_\sigma(i)$ if $\sigma_i$ is a call;
$\succ_\sigma^a(i)=i+1$ if $\sigma_i$ is not a call and $\sigma_{i+1}$ is not a return; and $\succ_\sigma^a(i)=\bot$ otherwise.

The \emph{past-successor} function $\succ_\sigma^-$ is  defined by $\succ_\sigma^-(i)=Q_{\sigma}(i)$.
\end{mydefinition}
\end{minipage}

\begin{mydefinition} The \emph{formulas} of CaRet are defined as follows:\\
\begin{math}
 \varphi\;::=\; p\mid\lnot\varphi\mid\varphi\land\varphi\mid\X^g\varphi\mid\varphi\U^g\varphi\mid\X^a\varphi\mid\varphi\U^a\varphi\mid\X^{-}\varphi\mid\varphi\U^{-}\varphi
\end{math}. \\
The superscript ``$g$'' comes from global, ``$a$'' from abstract, and ``$p$'' from past.
\end{mydefinition}

\begin{mydefinition}[semantics of CaRet]
Let $\sigma$ be  a finite word on $\widehat{2^{\AP}}=2^{\AP}\times\{\call,\internal,\ret\}$ and $i\in\mathbb{N}$ be such that $0\le i<|\sigma|$. We define the satisfaction relation $\models$ of CaRet formulas inductively as follows. 
\vspace{.2em}\\ 
\noindent
\begin{minipage}{\textwidth}
	\begin{small}
	$	\begin{array}{lll}
			\sigma,i\models p &\iff~& \sigma_{i}=(\tau,\place)\text{ for some $\tau$ such that } p\in \tau\\
			\sigma,i\models\lnot\varphi &\iff& \sigma,i\not\models\varphi\\
			\sigma,i\models\varphi\land\psi&\iff&\sigma,i\models\varphi\textrm{ and }\sigma,i\models\psi\\
			\sigma,i\models\X^g\varphi&\iff&\sigma,\succ^g_\sigma(i)\models\varphi, \textrm{ i.e., } \sigma,i+1\models\varphi\\
			\sigma,i\models\X^a\varphi&\iff&\succ^a_\sigma(i)\neq\bot\textrm{ and }\sigma,\succ^a_\sigma(n)\models\varphi \\
			\sigma,i\models\X^{-}\varphi&\iff&\succ^-_\sigma(i)\neq\bot\textrm{ and }\sigma,\succ^-_\sigma(n)\models\varphi\\
			\sigma,i\models\varphi\U^g\psi&\iff& \textrm{there is a sequence } i_0,i_1,\dots,i_k \;(\textrm{where } i_0=i) \textrm{ such that } \\ && \sigma,i_k\models\psi \textrm{ and } 
\forall j\in [0,k).
\;(i_{j+1}=\succ^g_\sigma(i_j) \textrm{ and } \sigma,i_j\models\varphi)\\
			\sigma,i\models\varphi\U^a\psi&\iff&\textrm{there is a sequence } i_0,i_1,\dots,i_k \;(\textrm{where } i_0=i) \textrm{ such that } \\ && \sigma,i_k\models\psi \textrm{ and } 
\forall j\in [0,k).
\;(i_{j+1}=\succ^a_\sigma(i_j) \textrm{ and } \sigma,i_j\models\varphi)\\
			\sigma,i\models\varphi\U^{-}\psi&\iff& \textrm{there is a sequence } i_0,i_1,\dots,i_k \;(\textrm{where } i_0=i) \textrm{ such that } \\ &&\sigma,i_k\models\psi \textrm{ and } 
\forall j\in [0,k).
\;(i_{j+1}=\succ^-_\sigma(i_j) \textrm{ and } \sigma,i_j\models\varphi)
		\end{array}
	$
\end{small}
\end{minipage}
\end{mydefinition}

\subsection{Automata, Regular and (Visibly) One-Counter}\label{subsec:prelimAutomata}

\begin{minipage}{\textwidth}
 \begin{mydefinition}[(regular) automata]\label{def:automata}
	A \emph{nondeterministic finite automaton} is a tuple $(Q,\Sigma,q_0,F,\Delta)$, where $Q$ is a finite set of states, $\Sigma$ is a finite alphabet, $q_0\in Q$ is an initial state, $F\subseteq Q$ is a set of final states, and $\Delta$ is a finite set of rules of the form $p\overset{a}{\to} q$, where $p,q\in Q$ and $a\in\Sigma$.
 \end{mydefinition}
\end{minipage}

\vspace{.2em}
For the two types of non-regular req-resp specs (RR3,4), we will use the following non-regular extensions of automata.
These notions are 
 proposed originally in \cite{AlurM04}; here  we follow the notations in \cite{Srba09}.

\vspace{.2em}
\noindent
\begin{minipage}{\textwidth}
\begin{mydefinition}[pushdown automata]
	A \emph{pushdown automaton} is a tuple $(Q,\Sigma,\Gamma,q_0,F,\Delta)$, where $Q$ is a finite set of states, $\Sigma$ is a finite alphabet, $\Gamma$ is a stack alphabet, $q_0\in Q$ is an initial state, $F\subseteq Q$ is a set of final states, and $\Delta$ is a finite set of rules of the form $pX\overset{a}{\to}q\alpha$, where $p,q\in Q$, $a\in \Sigma$, $X\in\Gamma$, and $\alpha\in\Gamma^*$. We assume that $Q$, $\Sigma$, $\Gamma$ are pairwise disjoint.
\end{mydefinition}
\end{minipage}

\vspace{.2em}
Pushdown automata manipulate the stack (by pushing and popping) as part of transition. For example, $pX\to q\mathit{XY}$ means pushing $Y$ to the stack, and $pX\to q\varepsilon$ means popping $X$ from the stack. The top of the stack is used to decide which transition to take.\vspace{.3em}

\noindent
\begin{minipage}{\textwidth}
\begin{mydefinition}[one-counter automata, cf.~\cite{Srba09}]
	A pushdown automaton is called \emph{one-counter} if the stack alphabet consists of two symbols, $\Gamma=\{I,Z\}$, and every rule is of the form $pI\overset{a}{\to} q\alpha$ or $pZ\overset{a}{\to} q\alpha Z$ with $\alpha\in\{I\}^*$. 
\end{mydefinition}
\end{minipage}

\vspace{.3em}
In this definition, every configuration reachable from $pZ$ is of the form $pI^nZ$. The symbol $I$ means adding one to the counter value, and $Z$ denotes the bottom of the stack, which means the counter value is zero. 

In depicting one-counter automata (cf.\ \cref{fig:automata_for_each_spec}), we express stack/counter manipulation accordingly: we write $c$ for a counter value, and  transition rules $pI\to q\mathit{II}$ and $pI\to q\varepsilon$ are expressed as $p\xrightarrow{c\,{:=}\,c+1}q$ and $p\xrightarrow{c\,{:=}\,c-1}q$.  A transition rule $pZ\to qZ$ means that if the counter value is zero, it is unchanged.




Pushdown automata are an expressive formalism that correspond to context-free languages, but this expressivity comes with increased complexity. The \emph{visible} restriction of pushdown automata~\cite{AlurM04}---and thus that of one-counter automata---aim to strike a middle-ground. In this restriction, each character $a\in \Sigma$ is either  \emph{call} (pushing one symbol to the stack),  \emph{return} (popping one symbol from the stack), or  \emph{internal} (not changing the stack size). Therefore the effect of $a\in \Sigma$ on the stack size  is fixed and independent of the current (control) state $p\in Q$ or the stack content. This restriction makes the class of recognized languages intersection-closed---unlike general context-free languages---and it has shown to be useful for various program verification problems (see e.g.,~\cite{AlurBE18,MurawskiW08}). 

In the following formal definition, note that the effect of $a\in \Sigma$ on the stack size can also depend on whether the stack is empty (i.e., whether $X=Z$). This is allowed in visible restrictions; see \cite[\S{}2.1]{AlurM04} and \cite[Rem.~2.2]{Srba09}.

\vspace{.2em}
\noindent
\begin{minipage}{\textwidth}
\begin{mydefinition}[visibly one-counter automata]\label{def:visiblyOneCounterAutom}
 Assume a partition
	 $\widehat\Sigma=\Sigma_c\cup\Sigma_{int}\cup\Sigma_r$ of the  alphabet $\widehat\Sigma$, much like \cref{def:matching-ret-innermost-call}. 
	A \emph{visibly one-counter automaton} is a one-counter automaton  $(Q,\widehat\Sigma,\Gamma,q_0,F,\Delta)$ that satisfies the following additional condition:  for each rule $pX\xrightarrow{a}q\alpha$, 1) $|\alpha|=2$ for each $a\in\Sigma_{c}$, 2) $|\alpha|=1$ for each  $a\in\Sigma_{int}$, and 3) $|\alpha|=0$ or $X=\alpha=Z$ for each $a\in\Sigma_{r}$. 
\end{mydefinition}
\end{minipage}


\section{A Decision Tree for Classifying Req-Resp Specs}\label{sec:clTree}

We propose a decision tree for classifying request-response specifications. We assume the satisfaction of $\G(\req\to\F\resp)$ as a baseline. 

\begin{figure}[tbp]
	\centering
	\includegraphics[width=\textwidth]{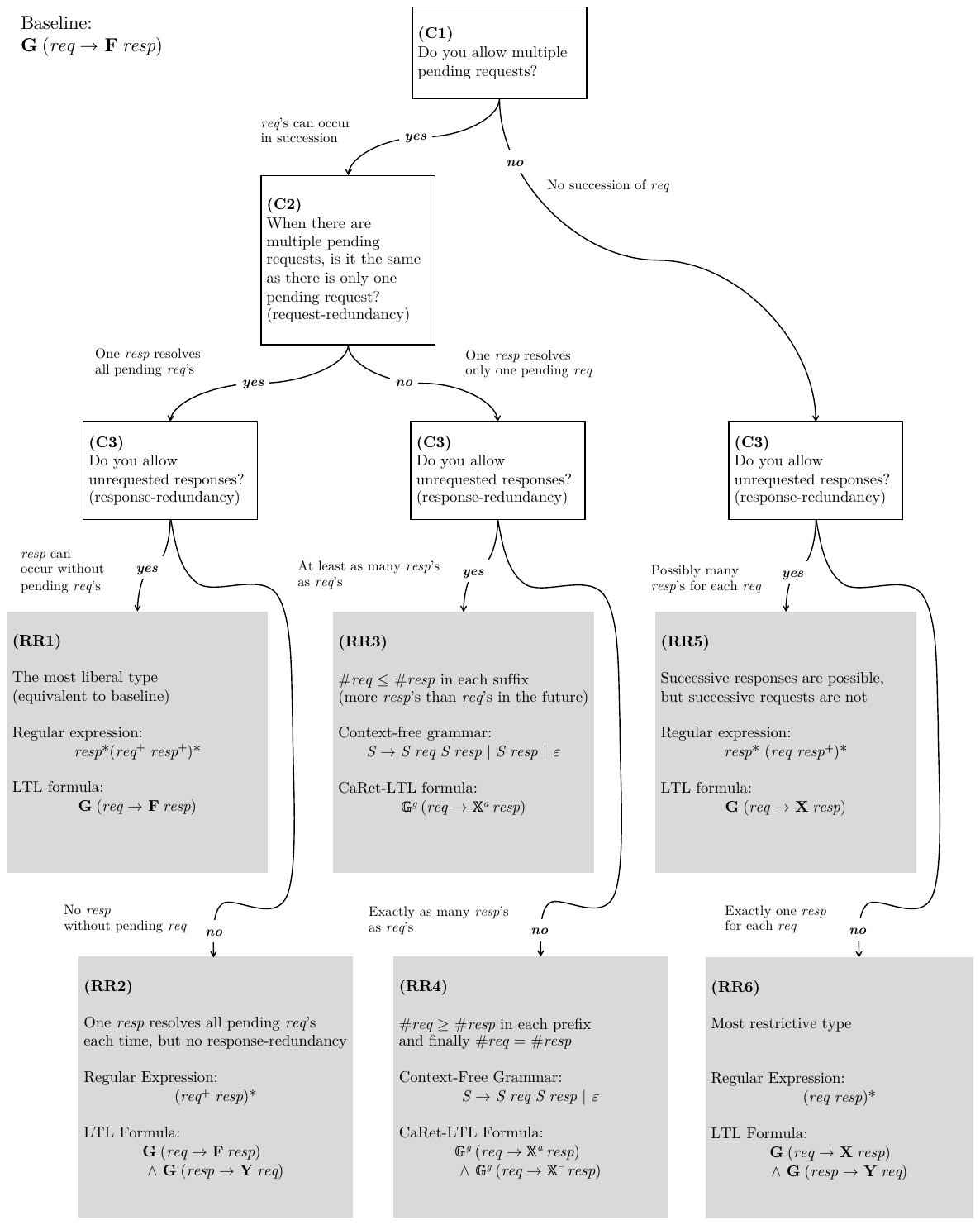}
	\caption{our proposed decision tree}
	\label{fig:clTree}
\end{figure}

Our proposal  is presented in \cref{fig:clTree}. The tree has three classification criteria (C1--C3), and they yield six types (RR1--RR6) of req-resp specifications. We explain and illustrate these types of specifications in this section, and we will provide mathematical formalizations of the six types in \cref{sec:formalization}.

\subsection{Three Classification Criteria}\label{subsec:criteria}

We identify three criteria (C1--C3) for classifying request-response specifications. Each criterion is expressed as a simple yes-no question. By answering them,  users can specify which type of specification is of interest to them.


\vspace{.3em}

\noindent
\begin{minipage}{\textwidth}
\begin{description}
	\item [(C1)] Do you allow multiple pending requests?
	\begin{itemize}
		\item Yes:  requests can occur in succession (e.g.,~$\req\;\req\;\resp$ is accepted).
		\item No: there is no succession of requests (e.g.,~$\req\;\req\;\resp$ is rejected).
	\end{itemize}
	\item [(C2)] (Request-redundancy) When there are multiple 
pending
requests, is it the same as there is only one 
pending request? (If yes, it means that those requests in the presence of a pending request are redundant.)
	\begin{itemize}
		\item Yes: one response resolves all 
pending
 requests at that moment (e.g.,~$\req\;\req\;\resp$ is identified with $\req\;\resp$).
		\item No: one response resolves only one 
pending request (e.g.,~when $\req\;\req\;\resp$, the last $\resp$ only resolves one of the two $\req$'s, and the remaining $\req$ is considered still pending).
	\end{itemize}
	\item [(C3)] (Response-redundancy) Do you allow unrequested responses? 
	\begin{itemize}
		\item Yes: extra responses may occur for a single request (e.g.,~$\req\;\resp\;\resp$ is accepted).
		\item No: all responses must be requested in advance (e.g.,~$\req\;\resp\;\resp$ is rejected, because the last $\resp$ is not requested).
	\end{itemize}
\end{description}
\end{minipage}


\begin{auxproof}
 Note that the meaning of the three criteria are not independent of each other. For example, the meaning of the term ``extra response'' in C3 is different depending on the answer to C2: in $\req\;\req\;\resp\;\resp$,
 \begin{itemize}
 \item the last $\resp$ is ``extra'' if the answer to C2 is yes (all the requests have been resolved by the last $\resp$), and
 \item the last $\resp$ is not ``extra'' if the answer to C2 is no.
 \end{itemize}
\end{auxproof}


\subsection{Our Decision Tree and Six Types of Req-Resp Specifications}\label{subsec:specifications}

These three criteria C1--C3 are combined to form a decision tree, shown in \cref{fig:clTree}. 
The tree first asks C1. If the answer is yes, it then asks C2. If not, there are only zero or one pending requests at any time---therefore there is no need to ask C2. The tree asks C3, at the end.
\begin{auxproof}
 , whose meaning differs depending on the answer to C2 (as we discussed above). 
\end{auxproof}
As a result, the tree yields six types (RR1--RR6) of request-response specifications.
Here are brief explanations of these six types. Their formalizations 
will be given in \cref{sec:formalization}; the classification 
is evaluated in \cref{sec:assessments}.

\underline{\bf RR1} is the most liberal out of the six types of req-resp specs, meaning that each of RR2--6 implies RR1. It also coincides with the \emph{baseline} specification ``every time $\req$ occurs,  $\resp$ will eventually follow,'' a spec that is commonly called \emph{the} request-response specification. RR1 allows both request-redundancy (C2) and response-redundancy (C3).


\underline{\bf RR2} restricts RR1 by removing response-redundancy: no $\resp$  allowed unless there is pending $\req$.  For example, RR1 holds for $\req\;\resp\;\resp$  but  not RR2. 


RR3--4 are the counterparts of RR1--2, respectively, where request-redundancy is disallowed. The latter means that, in presence of pending $\req$'s, additional  $\req$'s are no longer redundant; in other words, each $\resp$ resolves only one pending $\req$ (unlike in RR1--2 where $\resp$ resolves all pending $\req$'s). For RR3--4, therefore, we have to count the number of pending $\req$'s, making the specs non-regular. 


\underline{\bf RR3} is the version with response-redundancy, meaning that $\resp$'s are allowed even when there is no pending $\req$. We present the following mathematical characterization of RR3. Its proof is deferred to 
\FinalOrArxivVersion{\cite[Appendix~A]{Aiba_ICTAC2025_arxiv_extended_version}}{\cref{appendix:omittedProofs}}.

\vspace{.3em}
\noindent
\begin{minipage}{\textwidth}
\begin{mytheorem}[characterizing RR3]\label{thm:RR3}
Let $\sigma \in \{\req,\resp\}^{\ast}$ be a finite word. Then the following are equivalent. \vspace{.2em}

\noindent
\begin{minipage}{\textwidth}
\begin{enumerate}
 \item The word $\sigma$ satisfies RR3. Precisely, there is an injective correspondence $\rho$ from the index set $\{i\mid \sigma_{i}=\req\}$ to the  set $\{j\mid \sigma_{j}=\resp\}$ such that $i<\rho(i)$.
 \item In each suffix of $\sigma$, we have $\#\req\le\#\resp$ (i.e.,\ the number of $\req$'s is no bigger than that of $\resp$'s). \qed
\end{enumerate}
\end{minipage}
\end{mytheorem}
\end{minipage}\vspace{.3em}

\underline{\bf RR4} does not allow response-redundancy; accordingly, there must be a 1-1 correspondence $\rho$ from $\req$'s to $\resp$'s. Due to the baseline  $\G(\req\to\F\resp)$, $\rho$ should point forward ($i<\rho(i)$), too. We come to the following characterization.\vspace{.3em}

\noindent
\begin{minipage}{\textwidth}
\begin{mytheorem}[characterizing RR4]\label{thm:RR4}
 Let $\sigma \in \{\req,\resp\}^{\ast}$ be a finite word. Then the following are equivalent. \vspace{.3em}

\noindent
\begin{minipage}{\textwidth}
 \begin{enumerate}
 \item The word $\sigma$ satisfies RR4. Precisely, there is a bijective correspondence $\rho$ from the index set $\{i\mid \sigma_{i}=\req\}$ to the  set $\{j\mid \sigma_{j}=\resp\}$ such that $i<\rho(i)$.
 \item In each prefix of $\sigma$, we have $\#\req\ge\#\resp$. Moreover, we have $\#\req=\#\resp$ for the whole word.
 \item In each suffix of $\sigma$, we have $\#\req\le\#\resp$. Moreover, we have $\#\req=\#\resp$ for the whole word. \qed
 \end{enumerate}
\end{minipage}
\end{mytheorem}
\end{minipage}\vspace{.3em}


\underline{\bf RR5} allows successive responses, but no successive requests. Therefore it only  allows alternation of one request and a succession of responses. For example, RR5 holds for $\req\;\resp\;\resp$, but not for $\req\;\req\;\resp\;\resp$. 

\underline{\bf RR6} only allows one $\req$ and one $\resp$ to come alternately. RR6 holds for only those words of the form  $\req\;\resp\;\req\;\resp\dots\req\;\resp$.


\subsection{Examples of the Six Req-Resp Spec Types} \label{subsec:examples}
We present realistic examples of the six types RR1--6. There should be a vast variety of examples and we show only a fraction;  they should nevertheless serve the illustration purposes. They also show that all of RR1--6 have practical relevance.



\begin{figure}[tbp]\centering
	\begin{minipage}[t]{0.45\linewidth}
		\centering
		\includegraphics[width=0.78\textwidth,page=1]{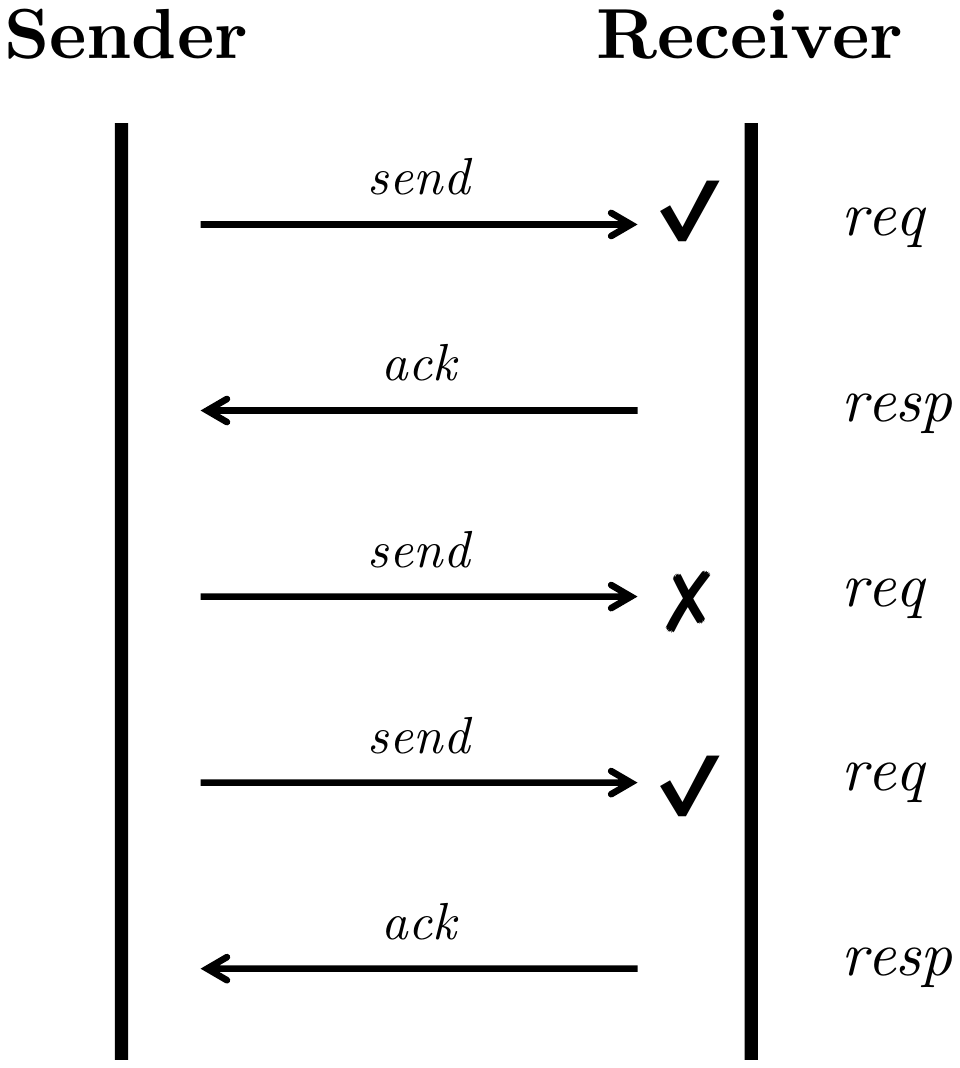}
		\subcaption{RR2: send and acknowledgment}
		\label{subfig:RR2}
	\end{minipage}\qquad
	\begin{minipage}[t]{0.35\linewidth}
		\centering
		\includegraphics[width=\textwidth,page=2]{fig/industrial_case_studies.pdf}
		\subcaption{RR3: MQTT QoS 1}
		\label{subfig:RR3}
	\end{minipage}
	\caption{network communication examples. \begin{small}\CheckmarkBold{}\end{small} is successful; \begin{small}\XSolidBrush{}\end{small} is not}
	\label{fig:casestudies}
\end{figure}

\underline{\bf RR1: Waiter}.  
Imagine calling a waiter at a restaurant. Here $\req$ is a call---by a hand sign, a bell, etc.---and $\resp$ is a waiter's attendance. It is possible (if not appreciated) to make multiple calls before being served (so C1 is yes); when a waiter comes then it resolves all the calls (C2 is yes; here we focus on a single table). Sometimes a waiter can come, too, without calls (thus C3 is yes).


\underline{\bf RR2: Send-Ack in Communication}.  
Consider
 the relationship between \emph{send} and \emph{acknowledgment} (Ack) in network communication (see \cref{subfig:RR2}).
Many communication protocols require explicit Acks for reliability. Here $\req$ is a message sent, and  $\resp$ is an Ack of successful delivery. A message can be sent many times before the sender gets Ack (so C1 is yes). Here we focus on single-threaded communication (no parallel sending of different messages), so one $\req$ resolves all $\resp$'s so far (C2 is yes).  Acks should not occur without sending (C3 is no).


\underline{\bf RR3: Broker in MQTT QoS 1}.  
\emph{MQTT} (message queuing telemetry transport) is a lightweight publish-subscribe network protocol  used e.g.,\ in IoT applications. MQTT features a network entity called a \emph{broker}; when an agent $A_{1}$  communicates a message to  $A_{2}$, $A_{1}$ (the \emph{publisher}) sends it to a broker, which sends it to $A_{2}$ (the \emph{subscriber}). Hence all messages get relayed by a broker.

MQTT offers three \emph{QoS levels} (quality of service) that give different guarantees. Among them, \emph{QoS 1} guarantees successful delivery of each message from a publisher to a subscriber. An example run of the protocol is in \cref{subfig:RR3}, where both the publisher and the broker sends a message repeatedly before its receipt is confirmed (by suitable ack messages, which we ignore in the current discussion). 

 We take the broker's view.  Let $\req$ stand for the broker receiving a message from the publisher, and $\resp$  for the broker sending a message to the subscriber (\cref{subfig:RR3}). This is  where use of RR3 is suitable: the broker can receive multiple messages before relaying them (C1 is yes); each received message ($\req$) must be relayed separately (C2 is no); and one $\req$ can trigger multiple $\resp$'s,  when some messages sent from the broker do not reach the subscriber (C3 is yes).

\underline{\bf RR4: Vending Machine}. Consider a vending machine. We let $\req$ stand for  accepting a coin (we assume that each item is one coin worth), and $\resp$ stand for dispensing a purchased item. This is a req-resp situation where we must have $\#\req=\#\resp$ in the end, so RR4 is suitable. Indeed, C1 is yes (here we assume that the machine allows pending purchases---otherwise we should use RR6), C2 is no (we must count the number of coins), and C3 is no (no free items).


\underline{\bf RR5: Reception with Numbered Tickets}.  
Consider a reception desk---or a bakery as in Lamport's bakery algorithm---that issues numbered tickets to clients. Here we take a client's viewpoint, and let $\req$ stand for the client taking a ticket, and $\resp$ stand for the client called by the desk. When the client has a ticket they will not take another (so C1 is no); the desk can call the client even if a ticket is not taken---e.g.,\ when there are no other clients or an emergent situation is obvious)---so C2 is yes. Therefore RR5 is a suitable spec type.

Note that, if we take a reception desk's viewpoint, RR3 is a suitable type, since the desk can issue multiple tickets to different clients. This example points to the following general principle: different agents in the same system have different viewpoints that make different spec types suited.


\underline{\bf RR6: Toggle Light Switch}. Consider a toggle light switch, where $\req$ is to turn on an electric light and and $\resp$ is to turn off. It is a req-resp situation---once a light is turned on, we want it to be off eventually. Since it is a toggle switch, $\req$ and $\resp$ must strictly alternate. 


\section{Formalization}\label{sec:formalization}

\noindent
\begin{minipage}{\textwidth}
We formalize the six types of req-resp specifications (RR1--RR6) introduced in \cref{sec:clTree}. We do so in three formalisms: 1) as  formal grammars, such as regular expressions and context-free grammars; 2) as logical formulas, in LTL or its variation; and 3) as automata, regular or visibly one-counter.
\end{minipage}


\subsection{Formalization as Grammars}\label{subsec:regexp_and_cfg}
Here we formalize each spec type as a formal grammar $G$. Then the monitoring problem---deciding a given word $w$ satisfies the spec or not---becomes the \emph{membership problem} in formal language theory, that is, to decide if $w\in L(G)$.


Let $\Sigma=\{\req,\resp\}$, and $w\in\Sigma^*$ be a finite word of $\req$'s and $\resp$'s.
Among the six types, the regular ones (RR1, RR2, RR5 and RR6) can be formalized as the following regular expressions. These should be intuitive.

\vspace{.2em}
\noindent
\begin{minipage}{\textwidth}
\begin{equation}\label{eq:regExpForRR1256} \small
 \begin{array}{llll}
  \text{\underline{\bf RR1}: }
&
\resp^*\,(\req^+\,\resp^+)^* 
\qquad\qquad\qquad
 &
  \text{\underline{\bf RR2}: }
 & 
(\req^+\,\resp)^*
\\
  \text{\underline{\bf RR5}: }
&
\resp^*\,(\req\,\resp^+)^*
 &
  \text{\underline{\bf RR6}: }
 & 
(\req\,\resp)^*
 \end{array}
\end{equation}
\end{minipage}


The remaining two (RR3--4) require counting  pending $\req$'s and thus are not regular. We thus use context-free grammars. They have only one nonterminal $S$. 

\vspace{.2em}
\noindent
\begin{minipage}{\textwidth}
\begin{equation}\label{eq:CFGForRR34}\small
\begin{array}{l}
   \text{\underline{\bf RR3}:  }
 S\to S\,\req\, S\,\resp\mid S\,\resp\mid\varepsilon
 \qquad\quad
  \text{\underline{\bf RR4}:  }
 S\to S\,\req\, S\,\resp\mid\varepsilon
\end{array}
\end{equation}
\end{minipage}\vspace{.2em}

\noindent Their correctness proofs, based on 
 \cref{thm:RR3,thm:RR4}, are not entirely trivial. 
 Proofs for RR3 (\cref{thm:suffixcount}) and RR4 (\cref{thm:prefixcount}) are deferred to 
 \FinalOrArxivVersion{\cite[Appendix~A]{Aiba_ICTAC2025_arxiv_extended_version}}{\cref{appendix:omittedProofs}}.


\vspace{.2em}
\noindent
\begin{minipage}{\textwidth}
\begin{mytheorem}[RR3 as a CFG]\label{thm:suffixcount}
The language	$\LRRthree=\{w\in\{a,b\}^*\mid\# a\leq \# b\,\text{ in each suffix of }w\}$ is produced by the rules $S\to SaSb\mid Sb\mid\varepsilon$. \qed
\end{mytheorem}
\end{minipage}

\vspace{.2em}
\noindent
\begin{minipage}{\textwidth}
\begin{mytheorem}[RR4 as a CFG]\label{thm:prefixcount}
$\LRRfour=\{w\in\{a,b\}^*\mid \# a\geq \# b \text{ in each }\linebreak \text{prefix of $w$, and $\# a= \# b$  in the whole $w$}\}$ is produced by $S\to SaSb\mid \varepsilon$. \qed
\end{mytheorem}
\end{minipage}

%
 The grammar characterization in~\cref{eq:CFGForRR34} trivially shows that RR4 implies RR3---the grammar for RR3 extends that for RR4 by an additional rule $S\to S\,\resp$. This implication is not trivial in the characterization in \cref{thm:RR3,thm:RR4}.




\begin{auxproof}
 \begin{figure}[tbp]
	\centering
	\begin{tabular}{c|c|c}
		\hline
		\small specifications & \small Reg.Exp/CFG&LTL/CaRet formulas\\\hline
		RR1 & $\resp^*(\req^+\resp^+)^*$&$\G(\req\to\F\resp)$\\
		RR2 & $(\req^+\resp)^*$&\;$\G(\resp\to\Y((\lnot\resp)\Sop\req))$\;\\
		RR3 & ~$S\to S\;\req\;S\;\resp\mid S\;\resp\mid\varepsilon$~~&$\G(\req\to\X^a\resp)$\\
		RR4 & $S\to S\;\req\;S\;\resp\mid\varepsilon$&$\G(\resp\to\X^-\req)$\\
		RR5 & $\resp^*(\req\;\resp^+)^*$&$\G(\req\to\X((\lnot\req)\U\resp))$\\
		RR6 & $(\req\;\resp)^*$&$\varphi_{\RR{2}}\land\varphi_{\RR{5}}$\\\hline
	\end{tabular}
	\caption{expressing each types of specification in Reg.Exp., CFG, and LTL}
	\label{fig:regexp_cfg_fml}
 \end{figure}
\end{auxproof}

\subsection{Formalization as Logical Formulas}\label{subsec:LTLformula}
Here we use another common formalism, namely temporal logics (specifically LTL (\cref{subsec:LTL}) and CaRet (\cref{subsec:caret})). Formalization in LTL makes specs amenable to many tools, such as model checkers (e.g.,~Spin and NuSMV) and monitoring tools (e.g.,~MonPoly). See \cref{sec:tools} for further discussion on monitoring tools.


\vspace{.2em}
\noindent
\begin{minipage}{\textwidth}
\begin{myassumption}[$\req,\resp$ as atomic propositions]\label{asmp:reqRespAtomicFormulas}
 Throughout the current \cref{subsec:LTLformula}, we set that $\req,\resp\in\AP$ are propositions. Moreover, we assume that \emph{exactly one of $\req$ and $\resp$ is true at each moment}. This removes such anomalies as $\req$ and $\resp$ occurring together, making the formalization compatible with the grammatical one in \cref{subsec:regexp_and_cfg} (where $\req, \resp\in \Sigma$ are characters). 
\end{myassumption}
\end{minipage}

\vspace{.2em}
\noindent
This assumption is posed mostly for the simplicity of presentation; see 
\FinalOrArxivVersion{\cite[Appendix~A.6]{Aiba_ICTAC2025_arxiv_extended_version}}{\cref{appendix:OnAsmpReqRespAtomicFormulas}} 
for technical justification. Lifting it should be strarightforward; doing so is future work.


As we discussed in \cref{sec:clTree}, we assume the LTL formula $\varphi_{\baseRR}=\G(\req\to\F\resp)$ as a baseline requirement for all of RR1--6. Therefore we are interested in  what additional requirement $\psi_{i}$ is needed to express each of RR1--6, which leads to a formalization $\varphi_{\RR{i}}=\varphi_{\baseRR}\land \psi_{i}$ of each type. Sometimes the additional requirement $\psi_{i}$ itself implies the baseline $\varphi_{\baseRR}$, in which case we do not see $\varphi_{\baseRR}$ explicitly in the formalization $\varphi_{\RR{i}}$.


\paragraph{Regular Spec Types}
We start with regular spec types. They are all expressible in LTL. Non-regular spec types (RR3--4) are formalized later in~\cref{eq:LTLFormalizationNonReg}.\vspace{.1em}

\noindent
\begin{minipage}{\textwidth}
\begin{equation}\label{eq:LTLFormalizationReg}\small
 \begin{array}{ll}
  \text{\underline{\bf RR1}: }\quad
&
\varphi_{\RR{1}}\;=\;
\G(\req\to\F\resp)
 \\
  \text{\underline{\bf RR2}: }
 & 
\varphi_{\RR{2}}\;=\;
\G(\req\to\F\resp)\land
\G(\resp\to\Y\req)
\qquad\qquad\qquad
\\
  \text{\underline{\bf RR5}: }
&
\varphi_{\RR{5}}\;=\;
\G(\req\to\X\resp)
\\
  \text{\underline{\bf RR6}: }
 & 
 \varphi_{\RR{6}}\;=\;
\G(\req\to\X\resp)\land
\G(\resp\to\Y\req)
\end{array}
\end{equation}
\end{minipage}\vspace{.1em}

\underline{\bf RR1} 
is the most liberal type, and it is equivalent to the baseline $\varphi_{\baseRR}$.


\underline{\bf RR2} poses an additional constraint that $\resp$ must not occur when there is no pending $\req$. In the current setting where 1) $\req$ and $\resp$ do not occur simultaneously and 2) either $\req$ or $\resp$ occurs at each moment (\cref{asmp:reqRespAtomicFormulas}), this translates to a simpler condition that ``each $\resp$ must be preceded by $\req$,'' that is, $\G(\resp\to\Y\req)$. Indeed, 
(Case 1)
 if $\G(\resp\to\Y\req)$ is true, then every $\resp$ is preceded by $\req$ that is pending, satisfying RR2;
(Case 2)
 if $\G(\resp\to\Y\req)$ is false, then there must be two consecutive $\resp$'s, violating RR2 (this contradicts with ``C3 is no''). 
 This additional constraint $\psi_{2}=\G(\resp\to\Y\req)$ does not imply the baseline ($\req\,\resp\,\req$ is a counterexample); so the formalization is $\varphi_{\RR{2}}=\varphi_{\baseRR}\land\psi_{2}$.

\vspace{.3em}
\noindent
\begin{minipage}{\textwidth}
 \begin{myremark}[RR2 without \cref{asmp:reqRespAtomicFormulas}]\label{rem:generalRR2}
 The above argument does not work in the general setting where \cref{asmp:reqRespAtomicFormulas} is not true.  We discuss  alternatives. 

 The formula $\psi=\F\resp\to((\lnot\resp)\U\req)$ is suggested in~\cite[p.85]{Goldblatt87} for an intention similar to RR2. It does not coincide with RR2, however, with a counterexample $\req\,\resp\,\resp$ (the formula is true but RR2 is false). 

 Its modification $\G\psi$ does not coincide with RR2 either. Consider the word $\req\, \emptyset\,\resp $, where $\emptyset$ means neither $\req$ or $\resp$ is true. RR2 holds for this word, but 
 $\psi
 $ is false at time 2, making $\G\psi$ false for the whole word.


 Another candidate is $\G(\resp \to\Y ((\lnot\resp) \Sop \req))$ suggested in~\cite{CimattiRS04}, and this seems to work. A proof of the coincidence with RR2 is left as future work.
 \end{myremark}
\end{minipage}

\begin{auxproof}
 RR2 imposes that the multiple $\req$ must be responded with a single $\resp$. A naive idea is $\G(\resp\to\Y\req)$, but this formula is not appropriate because we are considering on $2^{\{\req,\resp\}}$, permitting some situations where neither $\req$ nor $\resp$ comes. Another idea is $(\F \resp \to ((\lnot\resp) \U \req))$. This seems nice in appearance, also suggested in \cite[p.85]{Goldblatt87}, but this formula is not desirable, because it cannot deal with plural req-resp relationships. So, how about $\G(\F \resp \to ((\lnot\resp) \U \req))$? It turns out that this formula does not work well either; a counter example is shown in \cref{fig:counterexample_of_GF}, which indicates that a $\req$ corresponding to the nearest $\resp$ locates backward across the current state. In this situation, it is impossible to obtain a $\req$ that corresponds to a $\resp$; this is because $\U$ (until) operator refers only forward to the current state, so a $\req$ located in the past cannot be handled. 
 To make it work, $\G(\resp \to\Y ((\lnot\resp) \Sop \req))$ gives us the desired result. We use $\Sop$ to correspond to ``the most recent $\req$'' from the position of $\resp$. For your information, a similar formula is also found in \cite{CimattiRS04}.

 \begin{figure}[tbp]
	\centering
	\includegraphics[width=\textwidth]{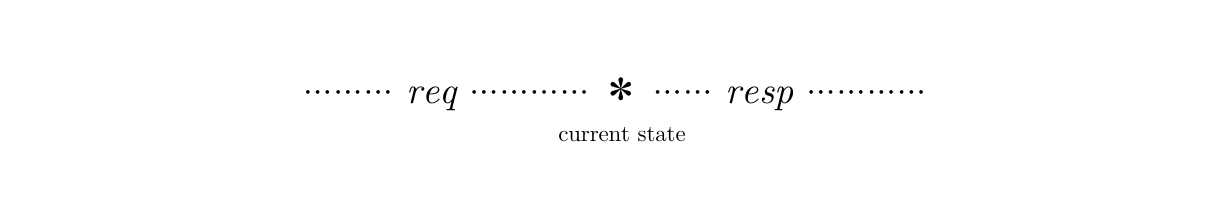}
	\caption{A Counterexample of $\G(\F\resp\to((\lnot\resp)\U\req))$}
	\label{fig:counterexample_of_GF}
 \end{figure}

 \[\varphi_{\RR{2}}=\G(\resp\to\Y(\lnot\resp\Sop\req))\]
\end{auxproof}

\underline{\bf RR5} adds, to the baseline $\varphi_{\baseRR}$, the constraint that $\req$'s cannot come in succession. Under \cref{asmp:reqRespAtomicFormulas}, this is the LTL formula $\G(\req\to\X\resp)$. Since this  implies $\varphi_{\baseRR}$, we obtain the formalization
\begin{math}
 \varphi_{\RR{5}}=
 \G(\req\to\X\resp)
\end{math}
 in~\cref{eq:LTLFormalizationReg}. 

\noindent
\begin{minipage}{\textwidth}
\begin{myremark}[RR5 without \cref{asmp:reqRespAtomicFormulas}]\label{rem:generalRR5}
We discuss, much like in \cref{rem:generalRR2}, a formula for RR5 in the general setting where \cref{asmp:reqRespAtomicFormulas} is not true. 

A good candidate is $\G(\req\to\X((\lnot\req)\U \resp))$, and one can show that this coincides with RR5, under its interpretation that $\resp$ occurring simultaneously with $\req$ does \emph{not} address $\req$. Further investigation is future work.
\end{myremark}
\end{minipage}

\vspace{.3em}
\underline{\bf RR6} is understood as the conjunction of RR2 and RR5. See~\cref{eq:LTLFormalizationReg}. Since RR5 implies the baseline $\G(\req\to\F\resp)$, we come to the formalization $\varphi_{\RR{6}}=\G(\req\to\X\resp)\land 
\G(\resp\to\Y\req)
$. Note that requiring $\G(\resp\to \X\req)$ is too strong---combined with $\G(\req\to\X\resp)$,   it does not allow a word to end.




\vspace{.8em}
\noindent
\begin{minipage}{\textwidth}
\paragraph{Non-Regular Spec Types}
For the non-regular spec types RR3--4, LTL is not enough, since all specs expressible in LTL are regular. We use a variant of CaRet (\cref{subsec:caret}), a minimal extension of LTL that accommodates the necessary kind of counting we need. In other words, using CaRet, we can  track injective/bijective correspondences between $\req$'s and $\resp$'s (cf.\ \cref{thm:RR3,thm:RR4}).
\end{minipage}

We first introduce the variant of CaRet we use; this is required since there is a subtle difference between the call-return correspondence that CaRet aims to capture, and the req-resp correspondence that we aim to capture. For distinction, we use symbols $\XX$, $\GG$, $\FF$, etc.\ for the temporal operators in  our variant of CaRet. 

This variant differs from the original CaRet (\cref{subsec:caret}) only in the definition of the innermost-call correspondence $Q_\sigma:\mathbb{N}\rightharpoonup\mathbb{N}$. Here, it is defined to be a partial function  that satisfies the following.

	\vspace{.1em}
	\noindent
	\begin{minipage}{\textwidth}
		\begin{itemize}
			\item If there is $j$ such that 1) $j<i$, 2) $\sigma_{j}$ is a call, and 3)  $R_\sigma(j)\ge i$ or $R_\sigma(j)=\bot$, then $Q_\sigma(i)$ is the greatest such $j$. 
			%
			\item Otherwise, $Q_\sigma(i)=\bot$.
		\end{itemize}
	\end{minipage}

\vspace{.1em}
The precise difference is in the condition 3) of the first item: here we have  $R_\sigma(j)\ge i$; in the original CaRet (\cref{subsec:caret}), we had  $R_\sigma(j)> i$.
This difference leads to the following property, making the logic suited for req-resp correspondence. 

\vspace{.2em}
\noindent
\begin{minipage}{\textwidth}
\begin{myproposition}\label{prop:QandRAreMutualInverseInModifiedCaret}
 In the above variant of CaRet, we have the following.
1) If $\sigma_{i}$ is a return and $Q_{\sigma}(i)$ is defined,  $R_{\sigma}(Q_{\sigma}(i))=i$ holds. 
 2) If $\sigma_{i}$ is a call and and $R_{\sigma}(i)$ is defined, $Q_{\sigma}(R_{\sigma}(i))=i$ holds.  \qed
\end{myproposition}
\end{minipage}

In this variant of CaRet, we also identify $\req$ with \emph{call} and $\resp$ with \emph{return}. Concretely, this means that in the extended alphabet $\widehat\Sigma=\Sigma\times\{\mathit{call},\mathit{int}, \mathit{ret}\}$, $\req$ is interpreted as $(\req,\mathit{call})$ and $\resp$ is  $(\resp,\mathit{ret})$.

Using this variant of CaRet, RR3--4 are naturally expressed as follows. \vspace{.1em}

\noindent
\begin{minipage}{\textwidth}
\begin{equation}\label{eq:LTLFormalizationNonReg}\small
  \begin{array}{ll}
  \text{\underline{\bf RR3}: }\quad
&
\varphi_{\RR{3}}\;=\;
\GG^{g}(\req\to\XX^{a}\resp)
 \\
  \text{\underline{\bf RR4}: }
 & 
\varphi_{\RR{4}}\;=\;
\GG^{g}(\req\to\XX^{a}\resp)\land
\GG^{g}(\resp\to\XX^{-}\req)
\end{array}
\end{equation}
\end{minipage}\vspace{.1em}

In $\varphi_{\RR{3}}$, by the definition of the semantics of $\XX^{a}$, one easily sees that $\XX^{a}\resp$ is equivalent to  $\XX^{a}\top$. We prefer the former presentation that is more intuitive. In $\varphi_{\RR{4}}$, the additional conjunct mandates a corresponding $\req$ to each $\resp$. 

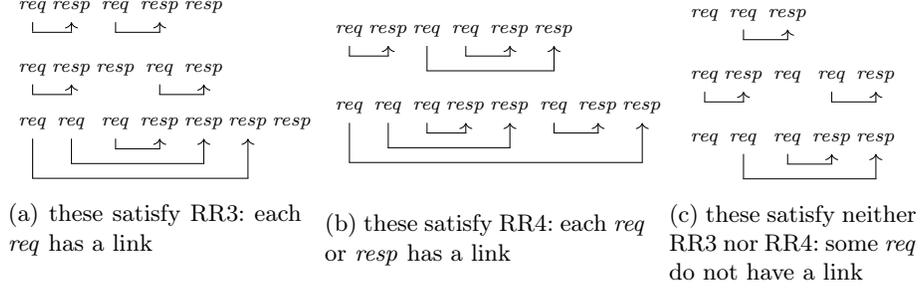
\begin{figure}[tbp]
	\centering
		\begin{minipage}[t]{0.32\linewidth}
			\centering
			\begin{tikzcd}[column sep=-6,row sep=3,style={font=\scriptsize}]
				\req \ar[d,dash,shorten >=-1pt]& \resp & \req \ar[d,dash,shorten >=-1.2pt] & \resp & \resp\\
				{} \ar[r,dash,shorten >=-4.3pt, shorten <=-4.3pt] & {} \ar[u,shorten <=-1pt] & {}\ar[r,dash,dash,shorten >=-4.3pt, shorten <=-4.3pt] & {}\ar[u,shorten <=-1.2pt] & {}\\
				\req \ar[d,dash,shorten >=-1.2pt] & \resp & \resp & \req \ar[d,dash,shorten >=-1.2pt] & \resp\\
				{} \ar[r,dash,shorten >=-4.3pt, shorten <=-4.3pt] & {}\ar[u,shorten <=-1.2pt] & {} & {}\ar[r,dash,shorten >=-4.3pt, shorten <=-4.3pt] & {}\ar[u,shorten <=-1.2pt]\\[-3pt]
				\req \ar[ddd,dash,shorten >=-1.2pt] & \req\ar[dd,dash,shorten >=-1.2pt] & \req \ar[d,dash,shorten >=-1.2pt] & \resp & \resp & \resp & \resp\\
				{} & {} & {}\ar[r,dash,shorten >=-4.3pt, shorten <=-4.3pt] & {}\ar[u,shorten <=-1.2pt] & {} & {} & {}\\[-4pt]
				{} & {} \ar[rrr,dash,shorten >=-4.3pt, shorten <=-4.3pt] & {} & {} & {} \ar[uu,shorten <=-1.2pt] & {} & {}\\[-4pt]
				{} \ar[rrrrr,dash,shorten >=-4.3pt, shorten <=-4.3pt] & {} & {} & {} & {} & {} \ar[uuu,shorten <=-1.2pt] & {}\\
			\end{tikzcd}
			\subcaption{these satisfy RR3: each $\req$ has a link}
		\end{minipage}
\;
		\begin{minipage}[t]{0.35\linewidth}
			\centering
			\begin{tikzcd}[column sep=-6,row sep=3,style={font=\scriptsize}]
				\req \ar[d,dash,shorten >=-1.2pt] & \resp & \req \ar[dd,dash,shorten >=-1.2pt] & \req \ar[d,dash,shorten >=-1.2pt] & \resp & \resp\\
				{} \ar[r,dash,shorten >=-4.3pt, shorten <=-4.3pt] & {}\ar[u,shorten <=-1.2pt] & {}  & {} \ar[r,dash,shorten >=-4.3pt, shorten <=-4.3pt] & {} \ar[u,shorten <=-1.2pt] & {}\\[-4pt]
				{} & {} & {}\ar[rrr,dash,shorten >=-4.3pt, shorten <=-4.3pt] & {} & {} & {}\ar[uu,shorten <=-1.2pt]\\ %
				\req \ar[ddd,dash,shorten >=-1.2pt] & \req \ar[dd,dash,shorten >=-1.2pt] & \req \ar[d,dash,shorten >=-1.2pt] &\resp &\resp & \req \ar[d,dash,shorten >=-1.2pt] & \resp & \resp\\
				{} & {} & {} \ar[r,dash,shorten >=-4.3pt, shorten <=-4.3pt] & {} \ar[u,shorten <=-1.2pt] & {} & {} \ar[r,dash,shorten >=-4.3pt, shorten <=-4.3pt] & {} \ar[u,shorten <=-1.2pt] & {}\\[-4pt]
				{} & {} \ar[rrr,dash,shorten >=-4.3pt, shorten <=-4.3pt] & {} & {} & {} \ar[uu,shorten <=-1.2pt] & {} & {} & {}\\[-4pt]
				{} \ar[rrrrrrr,dash,shorten >=-4.3pt, shorten <=-4.3pt] & {} & {} & {} & {} & {} & {} & {}\ar[uuu,shorten <=-1.2pt]
			\end{tikzcd}\vspace{4.5mm}
			\subcaption{these satisfy RR4: each $\req$ or $\resp$ has a link} 
		\end{minipage}
\;
	\begin{minipage}[t]{0.27\linewidth}
		\centering
		\begin{tikzcd}[column sep=-6,row sep=3,style={font=\scriptsize}]\\
			\req 
& \req \ar[d,dash,shorten >=-1.2pt] & \resp\\
 & {}\ar[r,dash,shorten >=-4.3pt, shorten <=-4.3pt] & {}\ar[u,shorten <=-1.2pt]\\ %
			\req \ar[d,dash,shorten >=-1.2pt] & \resp & 
\req 
& \req \ar[d,dash,shorten >=-1.2pt] & \resp\\
			{} \ar[r,dash,shorten >=-4.3pt, shorten <=-4.3pt] & {} \ar[u,shorten <=-1.2pt] &
 & {} \ar[r,dash,shorten >=-4.3pt, shorten <=-4.3pt] & {} \ar[u,shorten <=-1.2pt]\\ %
			\req
& \req \ar[dd,dash,shorten >=-1.2pt] & \req \ar[d,dash,shorten >=-1.2pt] & \resp & \resp\\
 & {} & {} \ar[r,dash,shorten >=-4.3pt, shorten <=-4.3pt] & {} \ar[u,shorten <=-1.2pt] & {}\\[-4pt]
			{} & {} \ar[rrr,dash,shorten >=-4.3pt, shorten <=-4.3pt] & {} & {} & {}\ar[uu,shorten <=-1.2pt]\\
		\end{tikzcd}
		\subcaption{these satisfy neither RR3 nor RR4: some $\req$ do not have a link}
	\end{minipage}
	\caption{examples for RR3--4.  The links indicate $R_{\sigma}$ (forward) and $Q_{\sigma}$ (backward)}
	\label{fig:matching_resp_RR3_and_RR4}
\end{figure}

Some (positive and negative) examples, with relevant links that indicate $R_{\sigma}$ and $Q_{\sigma}$ (that are mutually inverse in the sense of \cref{prop:QandRAreMutualInverseInModifiedCaret}), are shown in \cref{fig:matching_resp_RR3_and_RR4}.







\subsection{Formalization as Automata}\label{subsec:formalizationAsAutomata}

\begin{minipage}{\textwidth}
Here we present formalization of the six req-resp spec types as automata. Such representations as \emph{state machines} are  amenable to efficient algorithms: indeed, 
in the formal methods community and elsewhere,  grammars and logical formulas commonly get translated to automata  for algorithmic processing. 
\end{minipage}\vspace{.3em}

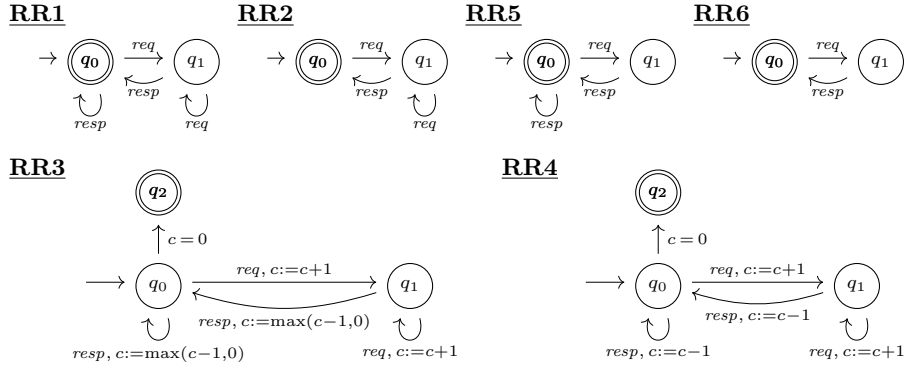
\begin{figure}[tbp] 
\begin{tabular}{llll}
\underline{\bf RR1} 
&
\underline{\bf RR2} 
 &
\underline{\bf RR5} 
& 
\underline{\bf RR6} 
\\
			\!\!\!\begin{tikzcd}[row sep=small,column sep=15,style={font=\scriptsize}]
				{}\ar[r,shorten <=8pt]& \doublecircled{$q_0$} \ar[r,"\mathit{req}"] \ar[loop below,"\mathit{resp}",looseness=4] & \circled{$q_1$} \ar[l,bend left,"\mathit{resp}"] \ar[loop below, "\mathit{req}",looseness=4]
			\end{tikzcd}
&
			\!\!\!\begin{tikzcd}[row sep=small,column sep=15,style={font=\scriptsize}]
				{}\ar[r,shorten <=8pt]& \doublecircled{$q_0$} \ar[r,"\mathit{req}"] & \circled{$q_1$} \ar[l,bend left,"\mathit{resp}"] \ar[loop below, "\mathit{req}",looseness=4]
			\end{tikzcd}
&
			\!\!\!\begin{tikzcd}[row sep=small,column sep=15,style={font=\scriptsize}]
				{}\ar[r,shorten <=8pt]& \doublecircled{$q_0$} \ar[r,"\req"] \ar[loop below,"\resp",looseness=4] & \circled{$q_1$} \ar[l,bend left,"\resp"] 
			\end{tikzcd}
&
			\!\!\!\begin{tikzcd}[row sep=small,column sep=15,style={font=\scriptsize}]
				{}\ar[r,shorten <=8pt]& \doublecircled{$q_0$} \ar[r,"\mathit{req}"]  & \circled{$q_1$} \ar[l,bend left,"\mathit{resp}"] 
			\end{tikzcd}
\end{tabular}\vspace{2mm}
\begin{tabular}{ll}
\underline{\bf RR3}
&
\underline{\bf RR4}
\\[-3mm]
 		\hspace{-13mm}	\begin{tikzcd}[row sep=4mm,column sep=70,style={font=\scriptsize}]
				{}&\doublecircled{$q_2$}\\
				{}\ar[r,shorten <=55pt]&\circled{$q_0$}\ar[r,"\req{,}\, c:=c+1"]\ar[u,"\,c\,=\,0\,"']\ar[loop below,"\resp{,}\, c:=\max(c-1{,}0)",looseness=4]&\circled{$q_1$}\ar[loop below,"\req{,}\,c:=c+1",looseness=4]\ar[l,bend left=6mm,"\resp{,}\, c:=\max(c-1{,}0)"]
			\end{tikzcd} \hspace{3mm}
&
		\hspace{-5mm}	\begin{tikzcd}[row sep=4mm,column sep=50,style={font=\scriptsize}]
				{}&\doublecircled{$q_2$}\\
				{}\ar[r,shorten <=35pt]&\circled{$q_0$}\ar[r,"\req{,}\,c:=c+1"]\ar[u,"\,c\,=\,0\,"']\ar[loop below,"\resp{,}\,c:=c-1",looseness=4]&\circled{$q_1$}\ar[loop below,"\req{,}\,c:=c+1",looseness=4]\ar[l,bend left=6mm,"\resp{,}\,c:=c-1"]
			\end{tikzcd}
\end{tabular}
	\caption{automata representation of the six types of req-resp specifications}
	\label{fig:automata_for_each_spec}
\end{figure}

The automata presentations are in \cref{fig:automata_for_each_spec}. Those for regular types (RR1,2,5,6) are usual automata (\cref{def:automata}); they happen to be deterministic as well. They can be easily obtained, e.g., by translating the regular expressions in \cref{subsec:regexp_and_cfg}. 

For non-regular types (RR3--4), we use visibly one-counter automata (\cref{subsec:prelimAutomata}). They are indeed \emph{visibly} one-counter, since each of $\req$ and $\resp$ has a fixed effect on the counter $c$ (incrementing and decrementing, respectively). The occurrence of  $\max$ in RR3 is allowed;
see \cref{def:visiblyOneCounterAutom} and the discussion before it.

\begin{auxproof}
 Although we can represent all the req-resp specifications using CFG or CaRet, these grammars and logics are highly expressive, and tools supporting such formalisms need a general and likely inefficient algorithm. In what follows, we demonstrate that all the req-resp specifications can be formulated using visibly one-counter DFAs, which is a mild extension of DFAs and a simple and efficient algorithm is sufficient for its monitoring.
\end{auxproof}

\section{Assessment}\label{sec:assessments} 

 We assess the validity of the classification
 in \cref{fig:clTree}. We do so by addressing three research questions: whether it is fine-grained enough (RQ1), not overly fine-grained (RQ2), and whether its organization as a tree is sensible (RQ3).


\vspace{.2em}
\myparagraph{RQ1: Is the Proposed Classification Sufficiently Discriminative?}
We argue it is. One reason is that the four regular types (RR1,2,5,6) cover all four possible patterns of singularity/plurality of $\req$'s and $\resp$'s. Specifically, we look at their regular expression presentations~\cref{eq:regExpForRR1256} and find, 
as  patterns therein, \vspace{.3em}

\noindent
\begin{minipage}{\textwidth}
\begin{small}
	\begin{center}
		$\req^+\,\resp^+\text{ in RR1,} \quad \req^+\,\resp \text{ in RR2,} \quad \req\,\resp^+ \text{ in RR5, ~ and } \req\,\resp \text{ in RR6}. $
	\end{center}
\end{small}
\end{minipage}\vspace{.3em}

\noindent These cover all four combinations of whether to put $+$ or not for $\req$ and $\resp$. 

In RR3--4 where we count the number of $\req$'s, the question is not about singularity/plurality, but more specifically about the numbers of $\req$'s and $\resp$'s. Because of the baseline spec  $\G(\req\to\F\resp)$, we must have $\#\req\le\#\resp$ for each suffix of a word $w$; this is RR3. RR4 imposes another condition, that is, $\#\req=\#\resp$ should hold for the whole $w$. Besides this, we have not found any reasonable additional condition so far.  Therefore the classification of this ``counting'' case into RR3 and RR4 is the best we can come up with now.

\vspace{.2em}
\myparagraph{RQ2: Is the Proposed Classification Not Overly Discriminative?}
The above arguments for RQ1 also show that our six types are distinguished for fundamental reasons (e.g.,~plurality of $\req$'s and $\resp$'s). The realistic examples in \cref{subsec:examples} demonstrate the need of this distinction, too: each of these examples demands the use of a specific type out of the six. 

One can also see that the three criteria C1--C3 in \cref{fig:clTree} are natural questions to ask when one fixes a req-resp specification. They do not seem overly detailed.

\vspace{.2em}
\myparagraph{RQ3: Is the Decision Tree Sensible?}
The three criteria C1--C3 are tightly coupled with the classification into RR1--6; so we can say that their validity, individually and collectively, has been addressed by the  above discussion. Therefore, here we focus on the specific \emph{order} of C1--C3 in \cref{fig:clTree}: is there any better order of C1--C3 when we ask them?


We argue, in fact, that the order in \cref{fig:clTree} is one that is \emph{logically derived}.  

Firstly, we need to ask C2 before we ask C3. This is because the meaning of C3 depends on the answer of C2.  More specifically, the meaning of ``unrequested responses'' changes depending on whether C2 is yes or no. For example, with the same word $\req\;\req\;\resp\;\resp$, if C2 is yes, then the last $\resp$ is considered to be unrequested; but if C2 is no, it should not be considered  unrequested.

Secondly, we need to ask C1 before we ask C2. This is because, if C1 is no, then there is no need to ask C2.

These two points requires the order C1-C2-C3, one that we have in \cref{fig:clTree}.





\section{Monitoring Tool Support, an Overview}
\label{sec:tools} 

In this section, we give an overview of available tools for monitoring the six types of req-resp specifications, with an emphasis  different ways of formalization (\cref{sec:formalization}). Here \emph{monitoring} of a spec means deciding if the spec is true in a given word $w$. 

\vspace{.2em}
\myparagraph{Monitoring: What  It Is and Why It Is Important}
When it comes to industrial usages of formal specs, monitoring is probably the most common task. Theoretically, monitoring can be seen as  a  special case of model checking: in monitoring, a system model is restricted to a single execution trace that is moreover finite; in contrast, in model checking, a system model is typically a state machine that can generate infinitely many execution traces each of which can be infinitely long. 
Consequently, \emph{fixed point reasoning}---that reduces infinitary to finitary---is a central theme in model checking, making it a theoretically intriguing topic. 

While fixed point reasoning is not much an issue in monitoring, monitoring does come with its own theoretical challenges, especially about (time and space) complexity. Specifically, an input word $w$ for monitoring can be very long (e.g., GBs of data), and monitoring such $w$ is only realistic if a monitoring algorithm has time complexity $O(|w|)$ and space complexity $O(1)$. Demand of practical speed is stronger in monitoring, too, since time constraints for monitoring tasks are often strict (especially when one conducts \emph{online} monitoring). 
Because of these differences, monitoring tools and model checking tools are usually separate. 

In practice, deployment of monitoring is considered easier than that of model checking. This is because the former requires only an execution trace $w$ of a target system---$w$ can be easily collected, e.g., from a log file---while the latter requires a white-box system model, building which can cost many person-months. Certainly, monitoring does not establish that  every behavior of a system is correct---it examines only one execution trace---but  its result often gives  useful insights.

Therefore  we focus on monitoring here, and present a current landscape of different tools available for different formalizations of req-resp specs. One can think of this overview  as an evaluation of different formalisms: some come with sophisticated tools; others do not.

\subsection{Tools Based on Grammars}\label{subsec:toolsGrammar}
\myparagraph{Regular Expressions} ``Monitoring'' regular expression specs is a special case of \emph{pattern matching}~\cite{Thompson68}. (Monitoring asks if the whole word matches a spec, while pattern matching usually answers all subwords that match.) 

There are many pattern matching tools; one  is the \texttt{grep} command of Unix.  Use of \texttt{grep} for monitoring  requires some care, however, such as the gap between characters and lines (\texttt{grep} is line-based) and the use of the \texttt{-o} option. 

Another class of tools is regex-matching engines, often offered as libraries in programming languages and frameworks. Examples are ones that come with Ruby, Google's RE2 (\url{https://github.com/google/re2}), to name a couple. 

We also discuss \emph{timed} extensions, i.e., formalisms that accommodate real-time constraints such as ``$\resp$ must come within 2.1 seconds,'' although this is outside the current paper's scope. \emph{Timed regular expressions}~\cite{AsarinCM02} are a well-known such extension; monitoring tools for them include MONAA~\cite{WagaHS18} and Montre~\cite{Ulus17}. 


\vspace{.2em}
\myparagraph{Context-Free Grammars (CFGs)} ``Monitoring'' CFGs largely amounts to \emph{parsing} (although the tree structure given by parsing is not needed). Parsing is a classic topic with a number of research works as well as tools. Examples of parsers are 
Lex/Yacc (available as Menhir in OCaml and flex/bison in C/C++), Antlr, and PEG-based parsers (such as Pyparsing in Python and Parsec in Haskell).

We tried Menhir, Antlr, and Pyparsing for our CFG formalization of RR3--4 (see~\cref{eq:CFGForRR34}). Menhir and Antlr worked well with both RR3--4. Pyparsing did not; this seems to be attributed to the absence of look-ahead in Pyparsing.

A timed extension of CFGs is proposed in \cite{SaeedloeiG10}. At this moment, its practical relevance in monitoring is unclear.

\subsection{Tools Based on Temporal Logics}
\myparagraph{LTL} LTL is probably the most common formalism for temporal specs. Its notable advantage is readability: unlike grammars and automata, LTL formulas allow natural language-like human interpretation (to a certain extent). 

Monitoring LTL specs is an active topic of research, with the work~\cite{DBLP:conf/cav/dAmorimR05} being one of the earliest. There are many tools for that, such as DejaVu~\cite{HavelundPU20}. 

Among those tools, MonPoly~\cite{BasinKZ17} stands out in its suitability to our current purpose. Firstly, it supports both future- and past-time operators in LTL, and we need both of them in our formalization (\cref{subsec:LTLformula}). DejaVu, in contrast, supports only past-time operators. Secondly, it has a track record of scalable real-world applications~\cite{Reger17}. Thirdly, it is an expressive tool and can accommodate real-time constraints as well as first-order specs. 

One possible disadvantage of MonPoly is that it is ``too expressive'' in a certain sense: sometimes a spec is deemed \emph{not monitorable} due to some theoretical constraints, and it is not easy to figure out what exactly is a problem. With our simple req-resp specs RR1,2,5,6, however, we did not encounter this problem.

\vspace{.2em}
\myparagraph{CaRet} CaRet seems to be a largely theoretical language, without much tool support or practical case studies. It is introduced in~\cite{AlurEM04}, and further theoretical studies are in~\cite{RosuCB08,DeckerLT13}. A monitor implementation is mentioned in~\cite{RosuCB08} but it does not seem to be available now. A timed extension of CaRet is not known, either.

\subsection{Tools Based on Automata}
\myparagraph{(Regular) Automata} Automata are a fundamental model of computation. ``Monitoring'' an automaton spec amounts to executing the automaton under a given word $w$; if $w$ is accepted, the spec is true for $w$. Executing an automaton is such a basic  task that there are numerous implementations. 

Automata are used in the backend of many tools which take specs in another formalism. For example, the work~\cite{DBLP:conf/cav/dAmorimR05} assumes a translation of  LTL formulas to  B\"uchi automata (e.g.,~\cite{Vardi95}). Most regex-matching engines (\cref{subsec:toolsGrammar}), too, first translate regular expressions to automata. 

\emph{Timed automata} are introduced in  \cite{AlurD94} as an extension of (regular) automata with real-time constraints. Monitoring timed automata specs is studied, e.g.,\ in~\cite{DBLP:phd/jp/Masaki20,DBLP:conf/rtss/AlurKV98}; MONAA~\cite{WagaHS18}  is a tool for monitoring timed automata (and timed regular expressions, too, via translation).

\vspace{.2em}
\myparagraph{Visibly Pushdown/One-Counter Automata}
 Visibly one-counter/pushdown automata, after their introduction in~\cite{AlurM04}, have been actively studied from the viewpoint of formal language theory~\cite{Srba09,AlurKMV05} as well as for application~\cite{AlurBE18,Schwentick07,MurawskiW08}.
However, their use for monitoring is rare. 

Regarding the timed extension, \emph{synchronized recursive timed automata} (SRTA)~\cite{UezatoM15} seem to be the closest. An SRTA extends a timed automaton with a stack.

\section{Conclusions and Future Work}

We studied a variety of request-response specifications, classifying them into six types  organized in a decision tree, and formalizing them in three different formalisms. The decision tree can be used for a practitioner to decide which type is suited for their application; they can do so by answering three questions (criteria C1--C3). Moreover, focusing on the monitoring task that is one of the most industry-relevant, we presented an overview of tool support.  
We believe all these will help practitioners when they use req-resp specs.

As theoretical  future work, we will investigate the general setting where \cref{asmp:reqRespAtomicFormulas} is lifted (see \cref{rem:generalRR2,rem:generalRR5}).
An applicational direction is to  develop further monitoring tools. For example, for non-regular RR3--4, the only tools available for practical use seem to be parsers (\cref{sec:tools}); dedicated monitoring tools for the current limited class of CFGs can be of practical interest.


\clearpage
\bibliographystyle{splncs04}
\bibliography{main}


\begin{ArxivBlock}

\clearpage
\appendix
\section{Omitted Details and Proofs}\label{appendix:omittedProofs}

\subsection{\cref{def:semLTL}, Abbreviations}\label{appendix:LTLsem}

\begin{mydefinition}[finite semantics of LTL]\label{def:semLTL}
	Let $\pi\in\left(2^{\mathrm{AP}}\right)^{\ast}$; it is called a finite structure of LTL. Let $\varphi,\psi$ be formulas, and $i$ be a natural number such that $0\leq i<|\pi|$. We inductively define the satisfaction relation $\models$ as follows. 
\begin{align*}
\begin{array}{l}
	\pi,i\models\top  \qquad
	\pi,i\models p \iff p\in\pi_{i}
	\\
	\pi,i\models\lnot\varphi\iff\pi,i\not\models\varphi \qquad
	\pi,i\models\varphi\land\psi\iff\pi,i\models\varphi\text{ and }\pi,i\models\psi
	\\
	\pi,i\models\X\varphi\iff i+1<|\pi|\text{ and }\pi,i+1\models\varphi
 	\\
	\pi,i\models\varphi\U\psi\iff\exists j\;\bigl( i\leq j <|\pi|\text{ and } \pi,j\models\psi\text{ and }\forall k\;(i\leq k<j\Longrightarrow \pi,k\models\varphi)\bigr)
 	\\
	\pi,i\models\Y\varphi\iff i>0\text{ and }\pi,i-1\models\varphi
 	\\
	\pi,i\models\varphi\Sop\psi\iff\exists j\;\bigl(0\leq j\leq i\text{ and }\pi,j\models\psi\text{ and }\forall k\;(j<k\leq i\Longrightarrow \pi,k\models\varphi)\bigr)\\
		\pi,i\models\varphi\R\psi\iff\forall j\;( i\leq j<|\pi|\implies(\pi,j\models\psi\textrm{ or }\exists k\;(i<k<j\textrm{ and }\pi,k\models\varphi)))\\
		\pi,i\models\F\varphi\iff\exists j\;(i\leq j <|\pi|\textrm{ and }\pi,j\models\varphi)\\
		\pi,i\models\G\varphi\iff\forall j\;(i\leq j<|\pi|\implies\;\pi,j\models\varphi)\\
		\pi,i\models\varphi\T\psi\iff\forall j\;(0\leq j\leq i\implies(\pi,j\models\psi\textrm{ or }\exists k\;(j<k\leq i\textrm{ and }\pi,k\models\varphi)))\\
		\pi,i\models\Oop\varphi\iff\exists j\;( 0\leq j\leq i\textrm{ and }\pi,j\models\varphi)\\
		\pi,i\models\Hop\varphi\iff\forall j\;(0\leq j\leq i\implies\pi,j\models\varphi)
	\end{array}
\end{align*}
\end{mydefinition}

\subsection{Proof of \cref{thm:RR3}}

\begin{myproof}{}
	[1 $\Rightarrow$ 2] 
   The injective correspondence $\rho$ establishes $\#\req\le\#\resp$ in the whole $\sigma$. Without loss of generality, we can drop all $\resp$'s that are not in the range of $\rho$; it suffices to show that Cond.~2 is true even after doing this.
   
   Since we have $i<\rho(i)$ for each $i$, each suffix of $\sigma$ has only the following three classes of characters: 1) $\req$  (whose corresponding $\resp$ via $\rho$ is necessarily in the same suffix), 2) $\resp$ that corresponds via $\rho$ to some $\req$ earlier in the suffix, and 3) $\resp$ corresponds to $\req$ outside the suffix. The first two classes induce a restriction of $\rho$ to the suffix. This establishes $\#\req\le\#\resp$ in that suffix.

   

   [2 $\Rightarrow$ 1] We define a required correspondence $\rho$ in the following inductive manner, maintaining  a counter $i$ (the current index) and a stack $\alpha$ of indices of pending requests. Initially $i=0$ and $\alpha=\varepsilon$ (empty). 
   \begin{itemize}
	\item If $\sigma_{i}=\req$, push $i$ to $\alpha$ (i.e., $\alpha:=i\alpha$).
	\item If $\sigma_{i}=\resp$ and $\alpha$ is nonempty, pop an element (say $j$) from $\alpha$ and set $\rho(j):=i$. If $\alpha$ is empty, do nothing.
	\item Update $i$ by $i:=i+1$.
   \end{itemize} 
   We argue that we obtain desired $\rho$; the only nontrivial part is that $\alpha$ becomes empty in the end, so that $\rho$ is defined everywhere over $\{i\mid \sigma_{i}=\req\}$. This follows from the fact that $|\alpha|+\#\req \le \#\resp$ is an invariant of the algorithm, where $\#\req$ and $\#\resp$ refer to the numbers in the suffix $\sigma_{i\le}$. (Checking this invariant requires Cond.~2.)
   Since $\#\resp$ becomes $0$ at the end, $\alpha$ also becomes empty.
   \qed
\end{myproof}

\subsection{Proof of \cref{thm:RR4}}

\begin{myproof}{}
  [1 $\Rightarrow$ 2] The bijective correspondence establishes $\#\req=\#\resp$ for the whole $\sigma$. Moreover, since $i<\rho(i)$, each $\resp$ in a prefix has its corresponding $\req$ in the same prefix. Therefore $\#\req\ge\#\resp$ in each prefix.

  [2 $\Rightarrow$ 1] We use the same construction of $\rho$ as in \cref{thm:RR3}. Cond.~2 here implies Cond.~2 of \cref{thm:RR3}, so we obtain $\rho$ that is injective. Moreover, Cond.~2 here yields that, whenever $\sigma_{i}=\resp$, the stack $\alpha$ is nonempty. Therefore no $\resp$ is wasted and $\rho$ is bijective. 

  [2 $\Leftrightarrow$ 3] Obvious from the fact that each suffix induces its ``complement'' prefix, and vice versa.
\qed
\end{myproof}

\subsection{Proof of \cref{thm:suffixcount}}

\begin{myproof} Let $G$ be the above CFG. We shall prove that $\LRRthree=L(G)$.

	($\subseteq$)\, By induction on the length of $w\in\LRRthree$. 
For the base case, the empty word $\varepsilon$ belongs to $\LRRthree$, and it is also produced by the CFG ($S\rightarrow\varepsilon$). 

For the step case, assume that $w\in \LRRthree$ and $|w|=k$. The last letter of $w$ must be $b$---otherwise $w\not\in \LRRthree$. We let $w=w'b$, with $|w'|=k-1$. We distinguish two cases. 


(Case 1) Assume that $\#a\leq\#b$ in each suffix of $w'$. In this case, by the induction hypothesis, we have $w'\in L(G)$ (i.e., $S\to^{\ast} w'$). We conclude $w=w'b\in L(G)$ by $S\to Sb \to^{\ast} w'b $.

(Case 2)
 Assume otherwise, that is, that we have $\#a>\#b$ in some suffix $w''$ of $w'$; we choose $w''$ to be the shortest such.
The suffix $w''$ must satisfy $\#a=\#b+1$, since $w''b$ is  a suffix of $w=w'b\in \LRRthree$. Moreover, $w''$ is of the form $w''=a w'''$---otherwise $w''=bw'''$ and $w'''$ has $\#a = \#b +2$, resulting in a suffix $w'''b$ of $w\in \LRRthree$ with $\#a = \#b +1$ (a contradiction). 

Overall, we have obtained a decomposition $w=w'''' a w''' b$ of $w$, and we have $\#a=\#b$ in $w'''$ and thus in $aw'''b$. 
Since $w\in \LRRthree$, we must have  $\#a\leq\#b$ in each suffix of $w''''$, yielding $w''''\in \LRRthree$. 

We further claim that $w'''\in \LRRthree$. Assume otherwise, then $w'''$ has a suffix $w'''''$ where $\#a>\#b$. Now $w'''''$ is a suffix of $w'=w'''' a w'''$, too, with $\#a>\#b$. Since $w'''''$, being a suffix of $w'''$, is shorter than $w''=aw'''$, this contradicts with the choice of $w''$ as the shortest one.

Therefore we have $w'''', w'''\in \LRRthree$, and  they are in $L(G)$ by the induction hypothesis. We then have a production 
\begin{math}
 S \to SaSb \to^{\ast} w'''' a w'''b =w
\end{math}.




($\supseteq$)
Straightforward by induction on production. \qed
\begin{auxproof}
 === Below: Daichi's original proof ===

	\begin{itemize}
		\item When $|w|=0$, the empty word $\varepsilon$ satisfies ($\Rightarrow$).
		\item Assume that ($\Rightarrow$) holds in the case of $|w|<k$ for $0<k$. Let $|w|=k$. 
		The last letter of $w$ must be $b$, by the assumption. Therefore we have $w=w'b$ for some $w'$ which is strictly shorter than $w$.
		\begin{itemize}
			\item If $\#a\leq\#b$ in each suffix of $w'$, then by the induction hypothesis, $w'$ is generated by the rules. The entire word $w$ is also generated by the rules, thanks to the rule $S \to Sb$.
			\item Otherwise, $\#a>\#b$ in some suffix of $w'$, so let us call it $w''$. The suffix $w''$ must satisfy $\#a=\#b+1$,\footnote{By the assumption on $w$, $\#a=\#b$ must hold in $w''b$.} and have $a$ as the first letter.\footnote{Note that $b$ cannot be the first letter of $w''$. If $w''$ starts with $b$, i.e., $w''=bu$ for some $u$, the suffix $u$ satisfies $\#a=\#b+2$. This means that $ub$, which is a suffix of the entire word $w$, satisfies $\#a=\#b+1$. This contradicts the assumption on $w$.} We have $w=w'''w''b$ for some $w'''$, and each suffix of $w'''$ satisfies $\#a\leq\#b$. Since $w''$ starts with $a$, we in fact have $w=w'''aw''''b$ for some $w''''$. Each suffix of $w''''$ satisfies $\#a\leq\#b$. Therefore, by the induction hypothesis on $w'''$ and $w''''$, these two words are generated by the rules. Thanks to the rule $S\to SaSb$, the entire word $w$ can be generated.
		\end{itemize}
	\end{itemize}
	($\Leftarrow$)\, By induction on the \emph{number of times} of rule-applications. (At each point, the contents of non-terminal symbols $S$ are considered to be undetermined, and we only count terminal symbols.)
	\begin{itemize}
		\item When the rule is applied once, we obtain $\varepsilon$, $SaSb$, or $Sb$, and all of them satisfy $\#a\leq\#b$ in each suffix.
		\item Let $w$ obtained through applying the rules $k-1$ times. Assume that ($\Leftarrow$) holds in case of less than $k$. After applying the rules $k$ times, we get either $w'=\dots\varepsilon\dots$, $w''=\dots SaSb\dots$, or $w'''=\dots Sb\dots$ from the word $w=\dots S\dots$. Then, by the induction hypothesis, the condition $\#a\leq\#b$ holds within each suffix.\qed
	\end{itemize} 
\end{auxproof}
\end{myproof}

\subsection{Proof of \cref{thm:prefixcount}}

\begin{myproof}
	Let $G$ be the CFG consisting of the rules $S\to SaSb\mid\varepsilon$. We shall prove that $\LRRfour=L(G)$.

	($\subseteq$) 
	The proof  method is much similar to \cref{thm:suffixcount}, because we also have $\#a\leq\#b$ in each suffix of $w$. The latter follows from the conditions that $\#a\geq\#b$ in each prefix of $w$ and that $\#a=\#b$ in the whole $w$. 

	We argue by induction on the length of $w\in\LRRfour$. The base case is exactly the same as \cref{thm:suffixcount}. For the step case, assume that $w\in\LRRfour$ and $|w|=k$. The last letter of $w$ must be $b$---otherwise $w\notin\LRRfour$ because $\#a=\#b$ does not hold in the whole $w$ when its prefix $w_{\leq k-1}$ satisfies the condition $\#a\geq\#b$. Let $w=w'b$, with $|w'|=k-1$. Note that $\#a=\#b+1$ in $w'$.

	(Case 1) Assume that $\#a\leq\#b$ in each suffix of $w'$.  It is clear that this assumption contradicts the condition $\#a=\#b+1$ in $w'$. 

	(Case 2) Now we are left with the case that we have $\#a>\#b$ in some suffix 
	of $w'$. 
	Through a discussion similar to \cref{thm:suffixcount}, we obtain a decomposition $w=w''''aw'''b$ of $w$. We have $\#a=\#b$ in both $w'''$ and $aw'''b$. Since $w\in\LRRfour$, we must have $\#a\geq\#b$ in each prefix of $w''''$, and $\#a=\#b$ in the whole $w''''$, yielding $w''''\in\LRRfour$. 
	We can also claim that $w'''\in\LRRfour$ in the same way as in \cref{thm:suffixcount}. Therefore we have $w'''',w'''\in\LRRfour$, and they are in $L(G)$ by the induction hypothesis. We then have a production $S\to SaSb\to^*w''''aw'''b=w$.

	($\supseteq$) Straightforward by induction on production.\qed
\end{myproof}

\begin{auxproof}
 An explicit proof, for our record:

...
\end{auxproof}

\subsection{On \cref{asmp:reqRespAtomicFormulas}}\label{appendix:OnAsmpReqRespAtomicFormulas}

	We fix $\AP = \{\req, \resp\}$. The finite-trace LTL semantics in \cref{appendix:LTLsem} is defined over $2^\AP$. By \cref{asmp:reqRespAtomicFormulas}, at each step exactly one of $\req$ and $\resp$ holds; hence neither $\emptyset$ nor $\{\req, \resp\}$ ever occurs as a letter of $2^\AP$. Accordingly, we work with the letters $\{\{\req\},\{\resp\}\}\subseteq 2^\AP$. For readability, we identify singleton letters with their atoms, and simply write $\req$ for $\{\req\}$ and $\resp$ for $\{\resp\}$.  


\subsection{Proof of \cref{prop:QandRAreMutualInverseInModifiedCaret}}
\begin{myproof}
	1) When $Q(i)\neq\bot$ (defined), $Q(i)=j$ such that a) $j<i$, b) $\sigma_j$ is a call, c) $R_\sigma(j)\geq i$, and d) $j$ is the greatest among such ones. Here we have $R_\sigma(j)=i$; otherwise if we assume $R(j)>j$, then there are the same numbers of calls and returns between $j$ and $R(j)$, which implies we can take another $j'$ that is greater than $j$, so this contradicts the fact that $j$ is the greatest one.

	2) When $R(i)\neq\bot$, $R(i)=j$ such that a) $j>i$, b) $\sigma$ is a return, c) the numbers of calls and returns between $i$ and $j$ are equal, and d) $j$ is the smallest among such ones. Then we have $Q(j)=i$, for i) $j>i$, ii) $\sigma_i$ is a call, iii) $R_\sigma(i)=j$, and moreover iv) $i$ is the greatest such ones; otherwise if we assume $i$ is not the greatest, then there exists $i'$ such that $i<i'<j$ and $R(i')=j$, which contradicts the fact that $j$ is the smallest one.  \qed
\end{myproof}

\end{ArxivBlock}


\end{document}